\documentclass[12pt]{article}

\usepackage[table,xcdraw]{xcolor} 
\usepackage{amsmath,amssymb}
\usepackage{graphicx}
\usepackage{array}
\usepackage{longtable}
\usepackage{booktabs}
\usepackage{colortbl}            
\usepackage{subcaption}
\usepackage{geometry}
\usepackage{etoolbox}            
\usepackage{orcidlink}

\usepackage[numbers,sort&compress]{natbib}

\usepackage{hyperref}            
\hypersetup{colorlinks=true, linkcolor=blue, citecolor=blue, urlcolor=blue}

\renewcommand{\arraystretch}{1.8}
\begin{document}
\baselineskip=20pt
\begin{center}
{\Large \textbf{$P$--$V$ criticality, Joule--Thomson expansion, and holographic heat engine of charged Hayward-AdS black holes with a cloud of strings and perfect fluid dark matter}}
\end{center}
\vspace{0.1cm}
\begin{center}
{\bf Ahmad Al-Badawi}\orcidlink{0000-0002-3127-3453}\\
Department of Physics, Al-Hussein Bin Talal University, 71111,
Ma'an, Jordan. \\
e-mail: ahmadbadawi@ahu.edu.jo\\
\vspace{0.1cm}
{\bf Faizuddin Ahmed}\orcidlink{0000-0003-2196-9622}\\Department of Physics, The Assam Royal Global University, Guwahati, 781035, Assam, India\\
e-mail: faizuddinahmed15@gmail.com\\
\vspace{0.1cm}
{\bf \.{I}zzet Sakall{\i}}\orcidlink{0000-0001-7827-9476}\\
Physics Department, Eastern Mediterranean University, Famagusta 99628, North Cyprus via Mersin 10, Turkey\\
e-mail: izzet.sakalli@emu.edu.tr (Corresponding author)
\end{center}
\vspace{0.1cm}

\begin{abstract}
We construct the charged Hayward-anti-de Sitter (AdS) black hole (BH) with a cloud of strings (CS) and perfect fluid dark matter (PFDM), and analyze its extended thermodynamic phase structure. The Hayward parameter $g$ replaces the central singularity with a de Sitter (dS) core, while the CS parameter $a$ and the PFDM parameter $\beta$ encode astrophysically motivated matter content. Treating the cosmological constant as pressure, we derive the thermodynamic quantities, verify the Smarr relation, and establish $P$--$V$ criticality with a van der Waals (vdW)-like small-large BH phase transition and mean-field critical exponents. The Gibbs free energy (GFE) exhibits the characteristic swallowtail below the critical pressure. The Joule-Thomson (JT) expansion yields $T_i^{\rm min}/T_c \approx 0.247$, roughly half the Reissner--Nordstr\"om-AdS value. The parameters $g$ and $Q$ contract the cooling region, $\beta$ expands it, and $a$ reshapes it non-monotonically. A holographic heat engine with a rectangular cycle gives efficiencies $\eta = 0.362$--$0.396$ and Carnot benchmarking ratios $\eta/\eta_C = 0.625$--$0.791$ across six configurations. The CS parameter improves the engine efficiency by reducing the enthalpy at fixed thermodynamic volume, while the PFDM parameter degrades it by adding gravitational enthalpy without contributing to the mechanical work.
\\[6pt]
\end{abstract}

{\color{black}
\section{Introduction} \label{isec1}

The discovery of Hawking radiation~\cite{isz01} and the formulation of BH thermodynamics by Bekenstein~\cite{isz02} revealed that BHs are not merely gravitational objects but genuine thermodynamic systems endowed with temperature, entropy, and free energy. In asymptotically AdS spacetimes, the thermodynamic description acquires additional structure due to the confining nature of the AdS boundary, which allows BHs to reach thermal equilibrium with their own radiation. The seminal work of Hawking and Page~\cite{isz03} demonstrated the existence of a first-order phase transition between thermal AdS and large Schwarzschild-AdS BHs, a result that was later reinterpreted in the context of the AdS/CFT correspondence~\cite{isz04} as a confinement-deconfinement transition in the dual gauge theory~\cite{isz05}. These connections between BH thermodynamics and quantum field theory have elevated the study of BH phase structure from a classical exercise to a central problem in theoretical physics \cite{Sucu:2025fwa,Gursel:2025wan,Aydiner:2025eii,Sucu:2025olo,Sucu:2025lqa}.

The extended phase space formalism, introduced by Kastor, Ray, and Traschen~\cite{isz06} and developed by Dolan~\cite{isz07,isz08} and Kubiz\v{n}\'ak and Mann~\cite{isz09}, promotes the cosmological constant to a thermodynamic variable identified with pressure through $P = -\Lambda/(8\pi)$, and the BH mass to enthalpy rather than internal energy. Within this formalism, the Reissner--Nordstr\"om-AdS (RN-AdS) BH exhibits a first-order small-large BH phase transition that is strikingly analogous to the liquid--gas transition of the van der Waals (vdW) fluid~\cite{isz09}, complete with matching critical exponents and a swallowtail structure in the GFE. This analogy has since been extended to a wide variety of AdS BHs, including Born--Infeld-AdS~\cite{isz10}, Gauss--Bonnet-AdS~\cite{isz11}, Lovelock-AdS~\cite{isz12}, and rotating BHs~\cite{isz13,isz14}, establishing $P$--$V$ criticality as a universal feature of charged and rotating AdS BH systems.

Beyond the static phase structure, the JT expansion provides a dynamical probe of BH thermodynamics. First studied for RN-AdS BHs by \"Okc\"u and Cebeci~\cite{isz15}, the JT expansion describes an isenthalpic process ($M = {\rm const}$) in which the BH depressurizes through a porous plug, and the sign of the JT coefficient $\mu_{\rm JT} = (\partial T/\partial P)_M$ determines whether the BH cools or heats. The inversion curve separating the two regimes has been computed for Kerr-AdS~\cite{isz16}, Gauss--Bonnet-AdS~\cite{isz17}, quintessence BHs~\cite{isz18}, and BHs in massive gravity~\cite{isz19}, revealing a characteristic suppression of the ratio $T_i^{\rm min}/T_c$ compared to ordinary fluids. Holographic heat engines, introduced by Johnson~\cite{isz20}, offer yet another window into the extended thermodynamics: a BH serves as the working substance of a thermodynamic cycle in the $P$--$V$ plane, and the engine efficiency encodes the competition among gravitational, electromagnetic, and matter degrees of freedom. Heat engine analyses have been performed for Born--Infeld-AdS~\cite{isz21}, Gauss--Bonnet-AdS~\cite{isz22}, and various modified gravity BHs~\cite{isz23,isz24}.

A separate line of development concerns the resolution of the central singularity in classical general relativity (GR). Bardeen~\cite{isz25} proposed the first regular BH solution, in which the Schwarzschild singularity is replaced by a dS core. Ay\'on-Beato and Garc\'{\i}a~\cite{isz26} later showed that the Bardeen metric can be interpreted as a gravitational field coupled to nonlinear electrodynamics (NED). Hayward~\cite{isz27} introduced a simpler regular BH metric characterized by a single parameter $g$ that controls the size of the dS core, with the metric function $f(r) = 1 - 2Mr^2/(r^3 + g^3)$. The Hayward solution has been widely studied for its thermodynamic properties~\cite{isz28,isz29}, quasinormal modes~\cite{isz30}, and gravitational lensing signatures~\cite{isz31}. The thermodynamics of Hayward-AdS BHs in the extended phase space was examined in~\cite{isz32,isz33}, where the vdW-like phase transition was confirmed and the effect of the regularity parameter on the critical point was characterized.

The incorporation of exotic matter sources into BH geometries provides additional physical motivation rooted in astrophysical observations. A CS, introduced by Letelier~\cite{isz34}, models a collection of one-dimensional objects threading the spacetime and modifies the metric through a deficit solid angle parametrized by $a$. BHs with CS backgrounds have been studied in the context of thermodynamics~\cite{isz35}, gravitational lensing~\cite{isz36}, and shadow observations~\cite{isz37}. PFDM, proposed by Kiselev~\cite{isz38} as a phenomenological model for the galactic dark matter halo, enters the metric through a logarithmic potential $\beta\ln(r/|\beta|)$ and has been coupled to various BH solutions~\cite{isz39,isz40}. The combination of regular BH geometries with CS and PFDM backgrounds is motivated by the desire to construct solutions that simultaneously address the singularity problem and incorporate astrophysically relevant matter content. Recent works have considered Bardeen-AdS BHs with PFDM~\cite{isz41} and Hayward BHs with CS~\cite{isz42}, but a full treatment combining all four parameters --- the Hayward regularity parameter $g$, the electric charge $Q$, the CS parameter $a$, and the PFDM parameter $\beta$ --- within the extended phase space formalism has not been carried out.

In this paper, we construct the charged Hayward-AdS BH in the simultaneous presence of a CS and PFDM and perform a detailed analysis of its extended thermodynamic phase structure. The metric function receives contributions from all four parameters, interpolating between a regular dS core at $r \to 0$ and an asymptotically AdS geometry at $r \to \infty$, with the singularity resolution guaranteed by the Hayward factor $(r^3 + g^3)^{-1}$ for all $g > 0$. Our aims are threefold. First, we derive the full set of thermodynamic quantities in the extended phase space, verify the Smarr relation, and map the $P$--$V$ critical behavior, including the critical point, the equation of state (EoS), and the GFE swallowtail structure. Second, we perform the JT expansion analysis and determine the inversion curves, the cooling-heating phase diagram, and the isenthalpic curves, paying particular attention to how the four parameters individually shape the cooling domain. Third, we construct a holographic heat engine with a rectangular cycle straddling the critical pressure and examine the engine efficiency, its parameter dependence, and its Carnot benchmarking across several limiting configurations (RN-AdS, Hayward+$Q$, Hayward+$Q$+CS, and the full model). A central question is how the regularity parameter, the CS tension, and the PFDM logarithmic potential compete or cooperate in determining the thermodynamic response of the BH across these different probes. We find, for instance, that the CS parameter improves the heat engine efficiency (by reducing the enthalpy at fixed thermodynamic volume) while the PFDM parameter degrades it (by adding gravitational enthalpy at the expansion stage), and that these two parameters have opposite effects on the JT cooling window as well.

The Event Horizon Telescope (EHT) observations of M87$^*$~\cite{isz43} and Sgr~A$^*$~\cite{isz44} have made regular BH models observationally testable, since the shadow size and photon ring structure depend on the near-horizon geometry that is modified by the regularity parameter~\cite{isz31,isz45}. While a direct shadow analysis is beyond the scope of the present work, the thermodynamic fingerprints we establish --- particularly the parameter-dependent shifts in the critical point, the inversion temperature, and the heat engine benchmarking ratio --- provide complementary constraints that connect to observational programs through the underlying metric parameters.

The paper is organized as follows. In Sec.~\ref{isec2}, we present the charged Hayward-AdS BH solution with CS and PFDM, analyze the metric function, and study the horizon structure including extremal configurations. In Sec.~\ref{isec3}, we derive the thermodynamic quantities in the extended phase space --- the Hawking temperature, the Bekenstein--Hawking entropy, the thermodynamic volume, the electric potential, and the conjugate variables to $a$, $\beta$, and $g$ --- and verify the extended first law and the Smarr relation. Section~\ref{isec4} is devoted to $P$--$V$ criticality: we compute the critical point, confirm the mean-field critical exponents $\alpha = 0$, $\tilde\beta = 1/2$, $\gamma = 1$, $\delta = 3$, analyze the GFE swallowtail, and map the coexistence curve in the $P$--$T$ plane. In Sec.~\ref{isec5}, we carry out the JT expansion, deriving the inversion curves, examining their dependence on $g$, $Q$, $a$, and $\beta$, and plotting the isenthalpic curves. Section~\ref{isec6} constructs the holographic heat engine with a rectangular cycle, studies the efficiency and its parameter dependence, and benchmarks the engine against the Carnot limit. We conclude in Sec.~\ref{isec7} with a summary of results and directions for future work.

\section{Charged Hayward-AdS BH with PFDM and CS} \label{isec2}

We consider a static, spherically symmetric spacetime described by the line element
\begin{equation}
ds^2 = f(r)\,dt^2 - \frac{dr^2}{f(r)} - r^2\left(d\theta^2 + \sin^2\theta\,d\phi^2\right),
\label{eq:metric}
\end{equation}
where the metric function for a charged Hayward-AdS BH immersed in PFDM and CS reads
\begin{equation}
f(r) = 1 - a - \frac{2Mr^2}{r^3 + g^3} + \frac{Q^2}{r^2} + \frac{\beta}{r}\ln\!\left(\frac{r}{|\beta|}\right) - \frac{\Lambda r^2}{3}\,.
\label{eq:fmetric}
\end{equation}
Here, $M$ denotes the ADM mass, $g$ is the Hayward regularization parameter that controls the dS core replacing the central singularity~\cite{Hayward2006}, $Q$ is the electric charge, $a$ is the CS parameter restricted to $0 \leq a < 1$~\cite{Letelier1979}, $\beta$ is the PFDM intensity parameter~\cite{Kiselev2003,Li2012}, and $\Lambda = -3/\ell^2 < 0$ is the cosmological constant with $\ell$ denoting the AdS curvature radius. The individual terms in~\eqref{eq:fmetric} encode distinct physical contributions: the CS parameter $a$ modifies the asymptotic value $f(r\to\infty) \to 1 - a$ by reducing the effective gravitational tension, the Hayward mass term replaces the Schwarzschild $2M/r$ singularity with a regular dS core of curvature $\sim 6M/g^3$ at $r \ll g$, the PFDM logarithmic term introduces a long-range gravitational drag that modifies the horizon structure at intermediate scales, and the AdS potential $-\Lambda r^2/3$ confines the geometry at large distances and enables the extended thermodynamic phase space.

An important structural feature of the Hayward regularization is the rational form of the mass function $2Mr^2/(r^3 + g^3)$. This stands in contrast to the Bardeen prescription $2Mr^2/(q^2 + r^2)^{3/2}$, where $q$ plays the role of a magnetic monopole charge~\cite{AyonBeato1999,AyonBeato2000}. Both models yield a regular dS core as $r\to 0$ and recover the Schwarzschild limit when the regularization parameter vanishes, yet they differ in the intermediate-$r$ regime. The Hayward denominator $r^3 + g^3$ produces a sharper transition from the dS core to the BH exterior compared to the Bardeen form $(q^2 + r^2)^{3/2}$, which leads to quantitatively different thermodynamic behavior, in particular near phase transitions and critical points. We will return to this comparison in Sec.~\ref{isec7}.

\subsection{Limiting cases} \label{isec2sub1}

The metric function~\eqref{eq:fmetric} encompasses several well-known BH geometries as special limits. Setting $g = Q = a = \beta = 0$ gives the Schwarzschild-AdS solution $f(r) = 1 - 2M/r - \Lambda r^2/3$. Restoring $Q \neq 0$ while keeping $g = a = \beta = 0$ yields the Reissner--Nordstr\"{o}m-AdS (RN-AdS) metric. The pure Hayward-AdS case corresponds to $Q = a = \beta = 0$. The Letelier-AdS solution~\cite{Letelier1979}, describing a BH threaded by a CS with no additional matter, follows from $g = Q = \beta = 0$, where $f(r) = 1 - a - 2M/r - \Lambda r^2/3$. Similarly, the PFDM-Schwarzschild-AdS solution~\cite{Kiselev2003,Li2012} arises when $g = Q = a = 0$. These limiting geometries allow us to isolate and compare the individual and combined effects of each parameter on the horizon structure and the thermodynamic quantities.

\subsection{Horizon structure} \label{isec2sub2}

The EH radii are determined by the condition $f(r_h) = 0$, which can be recast as
\begin{equation}
1 - a - \frac{2Mr_h^2}{r_h^3 + g^3} + \frac{Q^2}{r_h^2} + \frac{\beta}{r_h}\ln\!\left(\frac{r_h}{|\beta|}\right) - \frac{\Lambda r_h^2}{3} = 0\,.
\label{eq:horizon}
\end{equation}
Due to the transcendental nature of this equation --- stemming from the PFDM logarithmic contribution and the Hayward mass function --- an analytical solution is not attainable in closed form for generic parameter values. We therefore employ a numerical root-finding procedure. The results are collected in Table~\ref{tab:horizons}, which classifies the solutions into four distinct categories depending on the number and multiplicity of positive real roots of $f(r) = 0$.

When $g = Q = \beta = 0$ and $a = 0$, the Schwarzschild-AdS geometry admits a single EH at $r_h = 1.92830M$. Introducing the Hayward parameter $g > 0$ generates an inner (Cauchy) horizon $r_-$ in addition to the outer EH $r_+$, giving rise to a NE BH with two distinct horizons. For instance, the pure Hayward-AdS case with $g = 0.3$ yields $r_- = 0.11984M$ and $r_+ = 1.92147M$, where the inner horizon is controlled by the dS core radius while the outer horizon remains close to its Schwarzschild-AdS value. Similarly, the RN-AdS case ($Q = 0.5$) develops two horizons at $r_- = 0.13398M$ and $r_+ = 1.80274M$, with the charge-induced repulsion reducing $r_+$ more significantly than the Hayward regularization alone.

The CS parameter $a$ plays a distinct role: it raises the effective mass threshold for horizon formation, since $f(r) \to 1 - a$ at large $r$ rather than unity. As $a$ increases, the outer horizon radius grows --- the Letelier-AdS case ($a = 0.1$) has $r_h = 2.11683M$ compared to $1.92830M$ for Schwarzschild-AdS. When CS is combined with the Hayward and charge parameters ($g = 0.3$, $Q = 0.3$, $a = 0.1$), the outer horizon expands to $r_+ = 2.06875M$ while the inner horizon shifts to $r_- = 0.22643M$. The PFDM parameter $\beta$ modifies the horizon structure through its logarithmic potential. For $\beta > 0$, the contribution $(\beta/r)\ln(r/|\beta|)$ acts as a repulsive correction at small $r$ and an attractive one at large $r$, effectively compressing the horizon gap. Adding PFDM ($\beta = 0.5$) to the Hayward+$Q$ configuration drastically reduces the outer horizon to $r_+ = 1.35285M$.

As the parameters $g$, $Q$, $\beta$, and $a$ are varied, the inner and outer horizons approach each other. At a critical parameter combination, $r_-$ and $r_+$ merge into a single degenerate horizon satisfying $f(r_{\rm ext}) = 0$ and $f'(r_{\rm ext}) = 0$ simultaneously, corresponding to an Ext BH with vanishing surface gravity. Beyond this threshold, no real positive root exists and the metric describes a NS, as confirmed by the two entries at the bottom of Table~\ref{tab:horizons}.

\setlength{\tabcolsep}{10pt}
\renewcommand{\arraystretch}{1.6}
\begin{longtable}{|l|c|c|c|c|c|c|}
\hline
\rowcolor{orange!50}
\textbf{Configuration} & \textbf{$g/M$} & \textbf{$Q/M$} & \textbf{$a$} & \textbf{$\beta/M$} & \textbf{$r_-/M$} & \textbf{$r_+/M$} \\
\hline
\endfirsthead
\hline
\rowcolor{orange!50}
\textbf{Configuration} & \textbf{$g/M$} & \textbf{$Q/M$} & \textbf{$a$} & \textbf{$\beta/M$} & \textbf{$r_-/M$} & \textbf{$r_+/M$} \\
\hline
\endhead
\multicolumn{7}{|l|}{\cellcolor{gray!15}\textbf{Limiting cases}} \\
\hline
Schwarzschild-AdS & $0$ & $0$ & $0$ & $0$ & --- & $1.92830$ \\
\hline
Letelier-AdS & $0$ & $0$ & $0.10$ & $0$ & --- & $2.11683$ \\
\hline
Hayward-AdS & $0.30$ & $0$ & $0$ & $0$ & $0.11984$ & $1.92147$ \\
\hline
RN-AdS & $0$ & $0.50$ & $0$ & $0$ & $0.13398$ & $1.80274$ \\
\hline
\multicolumn{7}{|l|}{\cellcolor{gray!15}\textbf{Partial combinations}} \\
\hline
Hayward + $Q$ & $0.30$ & $0.30$ & $0$ & $0$ & $0.23002$ & $1.87774$ \\
\hline
Hayward + $Q$ + CS & $0.30$ & $0.30$ & $0.10$ & $0$ & $0.22643$ & $2.06875$ \\
\hline
Hayward + $Q$ + PFDM & $0.20$ & $0.40$ & $0$ & $0.50$ & $0.15193$ & $1.35285$ \\
\hline
\multicolumn{7}{|l|}{\cellcolor{gray!15}\textbf{Full model}} \\
\hline
Hayward + $Q$ + CS + PFDM (I) & $0.20$ & $0.40$ & $0.10$ & $0.50$ & $0.15056$ & $1.46358$ \\
\hline
Hayward + $Q$ + CS + PFDM (II) & $0.25$ & $0.50$ & $0.10$ & $0.80$ & $0.16036$ & $1.45446$ \\
\hline
Hayward + $Q$ + CS + PFDM (III) & $0.30$ & $0.50$ & $0.05$ & $0.60$ & $0.24669$ & $1.34956$ \\
\hline
Hayward + $Q$ + CS + PFDM (IV) & $0.35$ & $0.60$ & $0.15$ & $0.70$ & $0.31668$ & $1.40814$ \\
\hline
\multicolumn{7}{|l|}{\cellcolor{gray!15}\textbf{Naked singularity}} \\
\hline
NS (I) & $0.80$ & $1.20$ & $0$ & $0.30$ & --- & --- \\
\hline
NS (II) & $0.90$ & $1.50$ & $0.05$ & $0.40$ & --- & --- \\
\hline
\caption{Horizon radii for the charged Hayward-AdS BH with CS and PFDM at $M=1$ and $\Lambda=-0.03$. The configurations are grouped into limiting cases (single active parameter), partial combinations, the full model with all four parameters, and NS solutions where no real positive root of $f(r)=0$ exists. Single-horizon BHs have $r_- = $ ---, while NE BHs possess two distinct horizons $r_- < r_+$.}
\label{tab:horizons}
\end{longtable}

Solving the mass parameter from $f(r_h) = 0$ yields
\begin{equation}
M = \frac{r_h^3 + g^3}{2r_h^2}\left[1 - a + \frac{Q^2}{r_h^2} + \frac{\beta}{r_h}\ln\!\left(\frac{r_h}{|\beta|}\right) + \frac{r_h^2}{\ell^2}\right],
\label{eq:mass}
\end{equation}
where we used $\Lambda = -3/\ell^2$. This expression will serve as the starting point for deriving the Hawking temperature and the thermodynamic potentials in Sec.~\ref{isec3}.

\begin{figure}[ht!]
\centering
\includegraphics[width=0.95\textwidth]{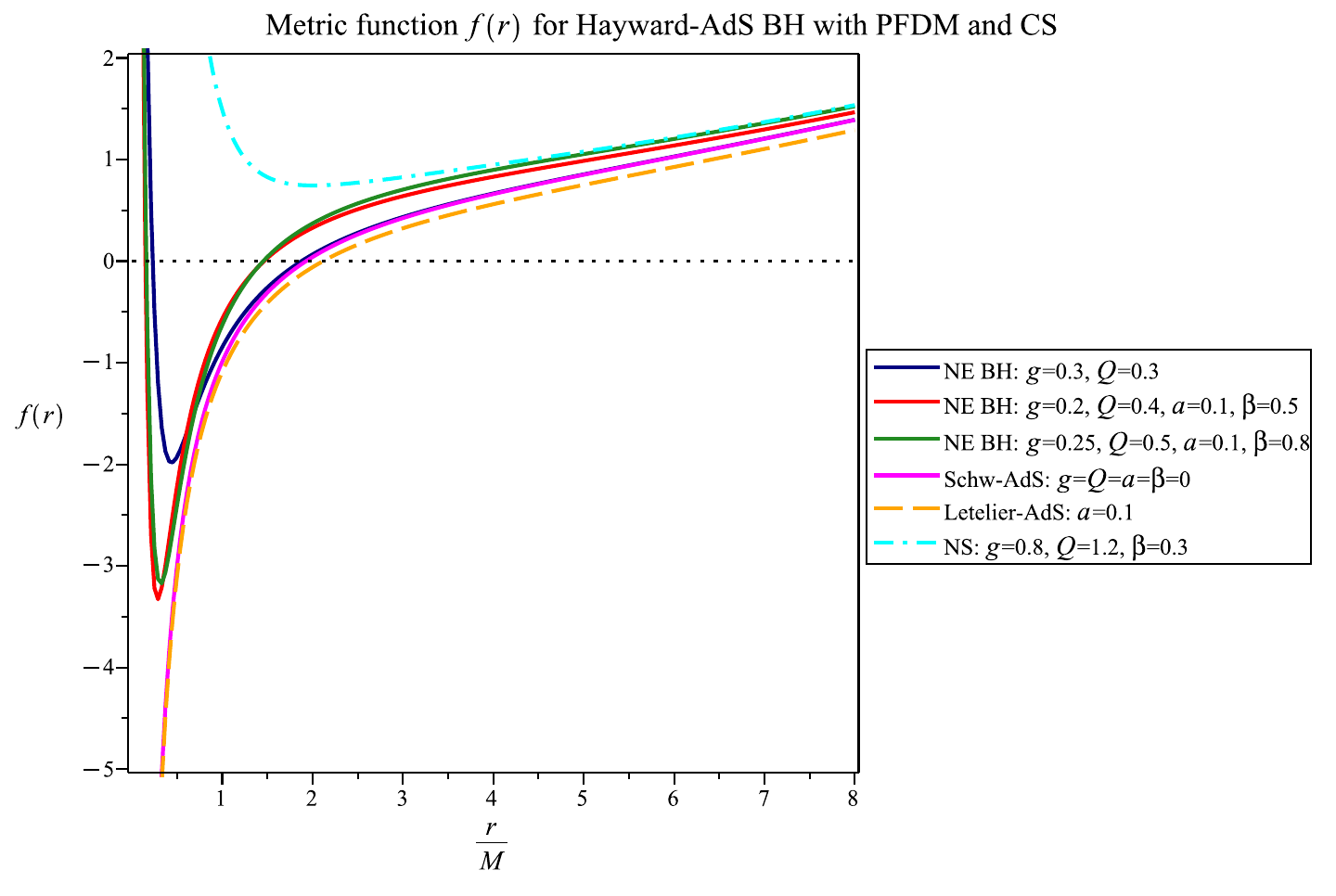}
\caption{Metric function $f(r)$ versus $r/M$ for the charged Hayward-AdS BH with CS and PFDM at $M=1$ and $\Lambda=-0.03$. The dotted horizontal line marks $f=0$. Navy, red, and green solid curves correspond to NE BH configurations with two horizons: $(g,Q)=(0.3,0.3)$, $(g,Q,a,\beta)=(0.2,0.4,0.1,0.5)$, and $(g,Q,a,\beta)=(0.25,0.5,0.1,0.8)$, respectively. The magenta curve is the Schwarzschild-AdS limit ($g=Q=a=\beta=0$), the dashed orange curve is the Letelier-AdS case ($a=0.1$), and the cyan dot-dashed curve is a NS configuration ($g=0.8$, $Q=1.2$, $\beta=0.3$) where $f(r)>0$ for all $r>0$.}
\label{fig:fmetric}
\end{figure}

The metric function $f(r)$ is plotted in Fig.~\ref{fig:fmetric} for representative parameter choices at fixed $M = 1$ and $\Lambda = -0.03$. Three NE BH configurations with two distinct horizons are shown: the Hayward+$Q$ case ($g = 0.3$, $Q = 0.3$) in navy, the full model case ($g = 0.2$, $Q = 0.4$, $a = 0.1$, $\beta = 0.5$) in red, and a second combined case ($g = 0.25$, $Q = 0.5$, $a = 0.1$, $\beta = 0.8$) in green. The Schwarzschild-AdS limit appears in magenta, and the Letelier-AdS case ($a = 0.1$) is shown as a dashed orange curve. The cyan dot-dashed curve corresponds to a NS configuration ($g = 0.8$, $Q = 1.2$, $\beta = 0.3$) where $f(r)$ remains positive for all $r > 0$. One observes that the Hayward regularization lifts $f(r)$ near the origin through the dS core, while the PFDM term and the CS parameter collectively reduce the outer horizon radius and narrow the gap between $r_-$ and $r_+$.

\begin{figure}[ht!]
\centering
\begin{subfigure}[b]{0.4\textwidth}
    \centering
    \includegraphics[width=\textwidth]{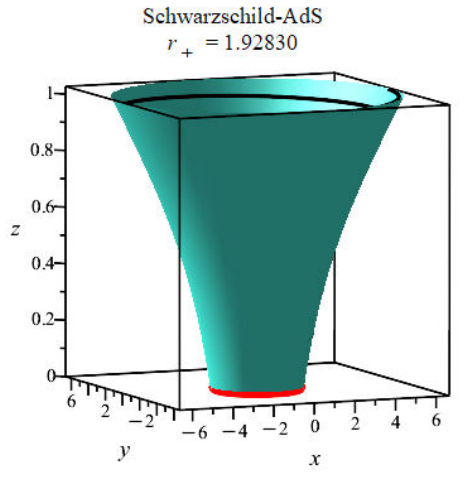}
    \caption{Schwarzschild-AdS: $g = Q = a = \beta = 0$.}
    \label{fig:3dv1}
\end{subfigure}
\hfill
\begin{subfigure}[b]{0.4\textwidth}
    \centering
    \includegraphics[width=\textwidth]{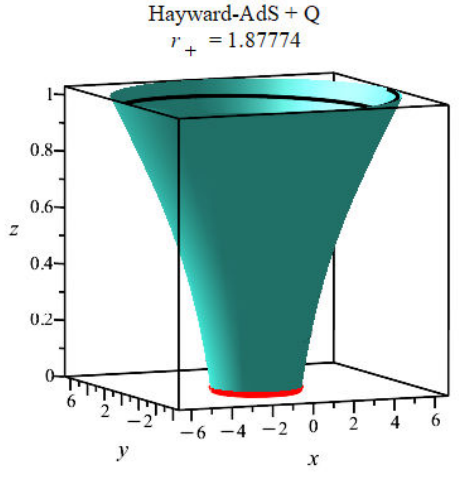}
    \caption{Hayward-AdS + $Q$: $g = 0.3$, $Q = 0.3$, $a = \beta = 0$.}
    \label{fig:3dv2}
\end{subfigure}

\vspace{0.5cm}

\begin{subfigure}[b]{0.4\textwidth}
    \centering
    \includegraphics[width=\textwidth]{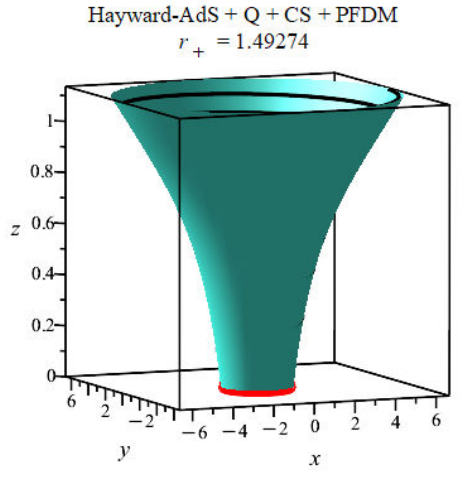}
    \caption{Hayward-AdS + $Q$ + CS + PFDM: $g = 0.3$, $Q = 0.3$, $a = 0.1$, $\beta = 0.5$.}
    \label{fig:3dv3}
\end{subfigure}
\hfill
\begin{subfigure}[b]{0.4\textwidth}
    \centering
    \includegraphics[width=\textwidth]{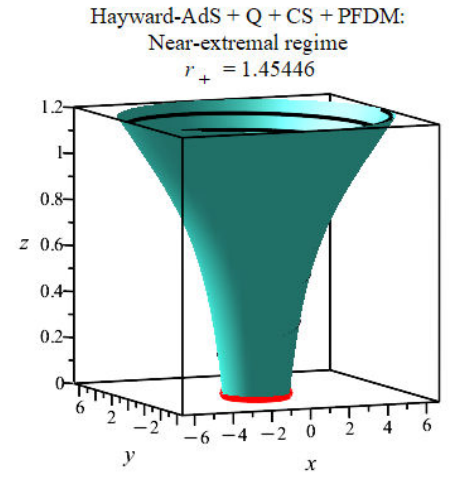}
    \caption{Near-extremal regime: $g = 0.25$, $Q = 0.5$, $a = 0.1$, $\beta = 0.8$.}
    \label{fig:3dv4}
\end{subfigure}
\caption{Three-dimensional embedding of the metric function $f(r)$ for the charged Hayward-AdS BH with CS and PFDM at $M=1$ and $\Lambda=-0.03$. The turquoise surface represents $f(r)$ over the $(x,y)$-plane, and the red ring marks the outer EH at $r = r_+$ where $f(r_+)=0$. From (a) to (d), the progressive narrowing of the throat reflects the combined effect of the Hayward regularization, electric charge, CS tension, and PFDM logarithmic potential in driving the BH toward extremality.}
\label{fig:3Dembedding}
\end{figure}

The three-dimensional embedding diagrams of the metric function are presented in Fig.~\ref{fig:3Dembedding} for four representative cases. The turquoise surface represents $f(r)$ over the $(x, y)$-plane with $x = r\cos\phi$ and $y = r\sin\phi$, and the red ring marks the outer EH at $r = r_+$ where $f(r_+) = 0$. The Schwarzschild-AdS case (Fig.~\ref{fig:3dv1}) exhibits the widest throat with $r_+ = 1.92830M$, reflecting the absence of any additional repulsive contributions. Introducing the Hayward parameter and charge (Fig.~\ref{fig:3dv2}) shrinks the throat to $r_+ = 1.87774M$, as the charge term $Q^2/r^2$ provides a repulsive barrier that reduces the horizon radius. When CS and PFDM are further included (Fig.~\ref{fig:3dv3}), the throat narrows considerably to $r_+ = 1.49274M$, demonstrating the combined effect of the string cloud tension and the dark matter logarithmic potential in reducing the gravitational well. The near-extremal regime (Fig.~\ref{fig:3dv4}) exhibits the narrowest throat with $r_+ = 1.45446M$, where the inner and outer horizons are approaching coalescence and the surface gravity is close to vanishing. The progressive narrowing of the embedding surface from Figs.~\ref{fig:3dv1} to~\ref{fig:3dv4} provides a geometric visualization of how the interplay among $g$, $Q$, $a$, and $\beta$ drives the BH toward extremality.

\section{Thermodynamic quantities in the extended phase space} \label{isec3}

In the extended phase space formalism~\cite{Kastor2009,Dolan2011a,Dolan2011b,Kubiznak2012}, the cosmological constant is promoted to a thermodynamic variable identified with the pressure
\begin{equation}
P = -\frac{\Lambda}{8\pi} = \frac{3}{8\pi\ell^2}\,,
\label{eq:pressure}
\end{equation}
and the BH mass $M$ is interpreted as enthalpy $M \equiv H$ rather than internal energy. This identification is dictated by the scaling behavior of the Komar integral and ensures consistency of the Euler relation for AdS BHs. The conjugate thermodynamic volume $V$ then enters the extended first law through the $VdP$ term. For the charged Hayward-AdS BH with CS and PFDM described by~\eqref{eq:fmetric}, the full set of thermodynamic quantities is derived below.

\subsection{Hawking temperature} \label{isec3sub1}

The Hawking temperature follows from the surface gravity at the outer EH $r_+$ via $T = f'(r_+)/(4\pi)$. Differentiating~\eqref{eq:fmetric} with respect to $r$ and evaluating at $r = r_+$ where $f(r_+) = 0$ yields
\begin{equation}
T = \frac{1}{4\pi}\left[\frac{2Mr_+(2r_+^3 - g^3)}{(r_+^3 + g^3)^2} - \frac{2Q^2}{r_+^3} + \frac{\beta}{r_+^2}\left(1 - \ln\frac{r_+}{|\beta|}\right) + \frac{2r_+}{\ell^2}\right].
\label{eq:temperature_raw}
\end{equation}
It is more useful to eliminate $M$ in favor of the horizon data using~\eqref{eq:mass}, yielding an expression for $T$ in terms of $r_+$ and the parameters $(g, Q, a, \beta, P)$ alone. In the limit $g = Q = a = \beta = 0$, this reduces to the Schwarzschild-AdS result $T = (1 + 8\pi P r_+^2)/(4\pi r_+)$, while setting only $g = a = \beta = 0$ recovers the RN-AdS temperature $T = (1 - Q^2/r_+^2 + 8\pi P r_+^2)/(4\pi r_+)$.

\begin{figure}[ht!]
\centering
\includegraphics[width=0.75\textwidth]{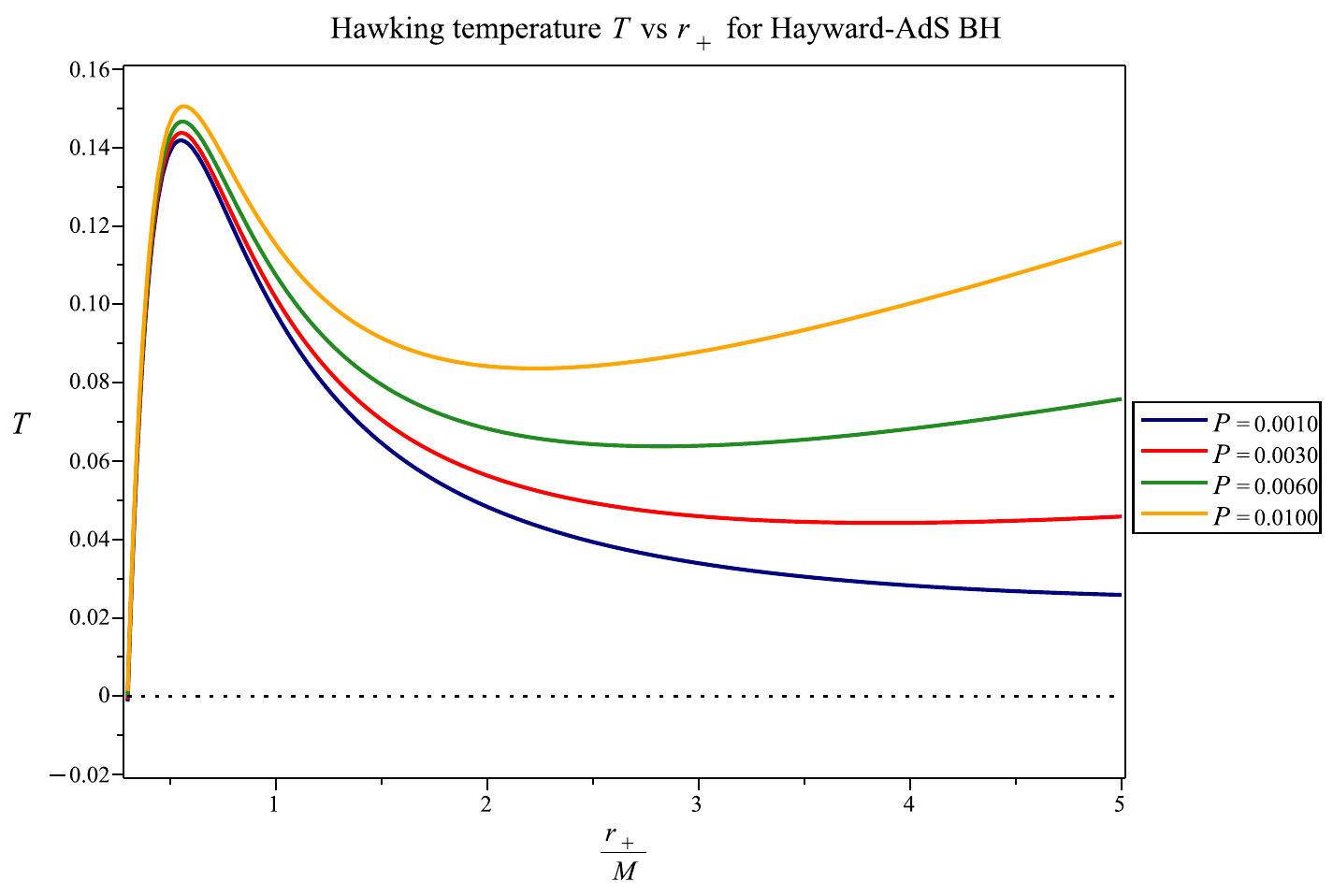}
\caption{Hawking temperature $T$ versus outer horizon radius $r_+/M$ for the charged Hayward-AdS BH with CS and PFDM at $g = 0.3$, $Q = 0.3$, $a = 0.1$, $\beta = 0.5$. Different curves correspond to $P = 0.0010$ (navy), $P = 0.0030$ (red), $P = 0.0060$ (green), and $P = 0.0100$ (orange). All curves share a local maximum near $r_+ \approx 0.6M$, while the large-$r_+$ behavior is controlled by the AdS pressure term $\sim 2Pr_+$.}
\label{fig:T_vs_P}
\end{figure}

\begin{figure}[ht!]
\centering
\begin{subfigure}[b]{0.48\textwidth}
    \centering
    \includegraphics[width=\textwidth]{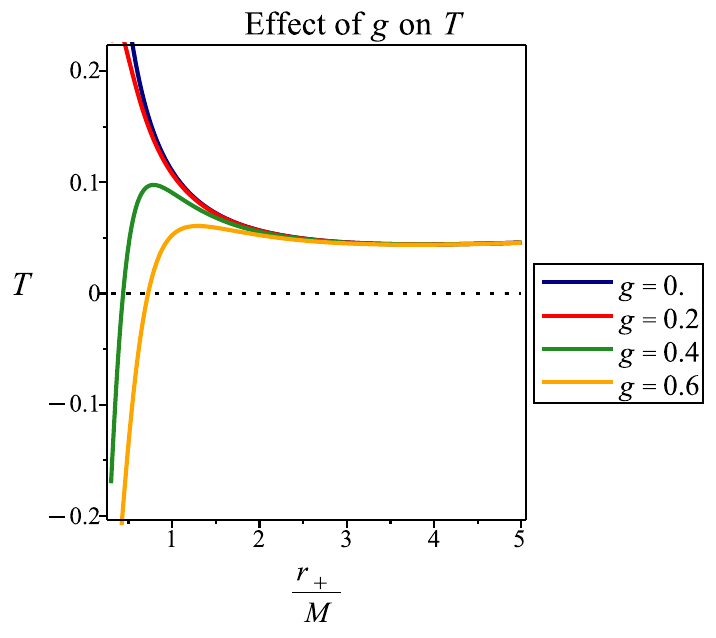}
    \caption{Effect of $g$ at $Q = 0.3$, $a = 0.1$, $\beta = 0.5$.}
    \label{fig:T_g}
\end{subfigure}
\hfill
\begin{subfigure}[b]{0.48\textwidth}
    \centering
    \includegraphics[width=\textwidth]{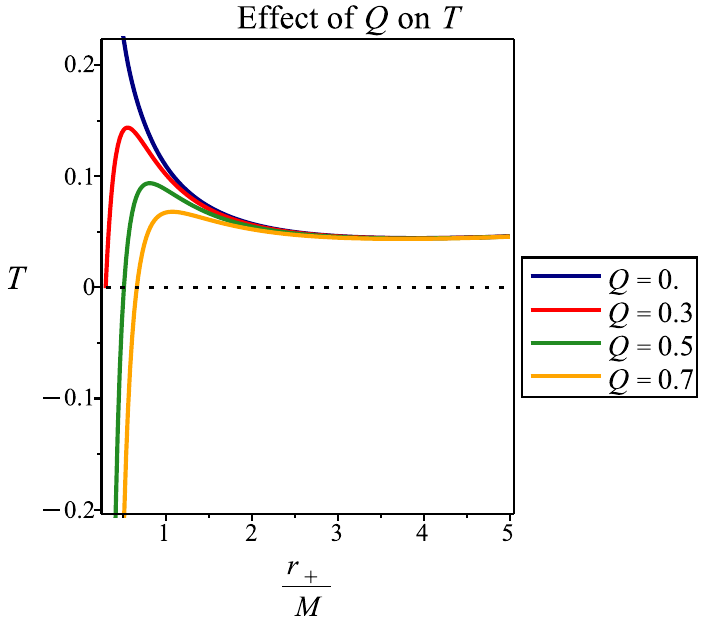}
    \caption{Effect of $Q$ at $g = 0.3$, $a = 0.1$, $\beta = 0.5$.}
    \label{fig:T_Q}
\end{subfigure}

\vspace{0.5cm}

\begin{subfigure}[b]{0.48\textwidth}
    \centering
    \includegraphics[width=\textwidth]{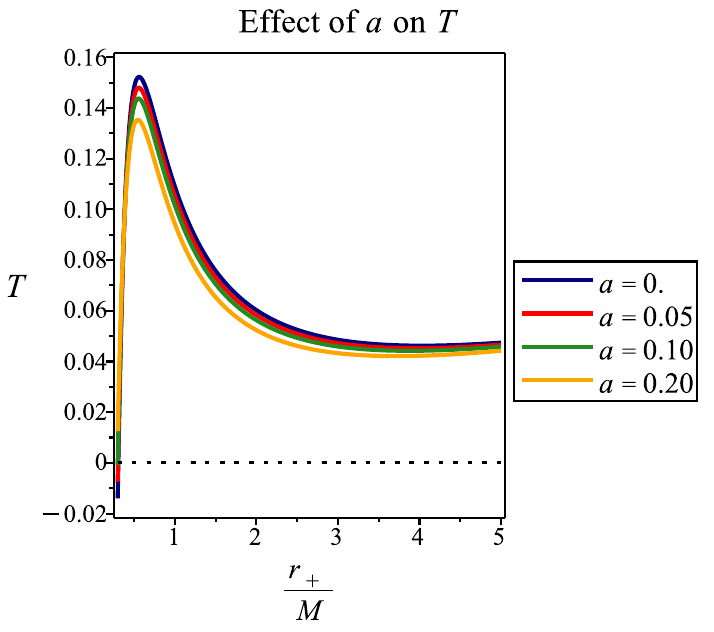}
    \caption{Effect of $a$ at $g = 0.3$, $Q = 0.3$, $\beta = 0.5$.}
    \label{fig:T_a}
\end{subfigure}
\hfill
\begin{subfigure}[b]{0.48\textwidth}
    \centering
    \includegraphics[width=\textwidth]{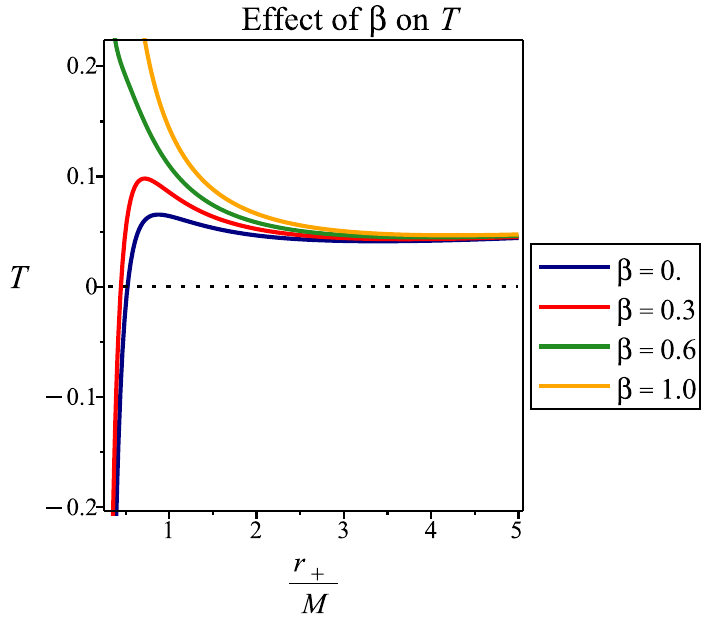}
    \caption{Effect of $\beta$ at $g = 0.3$, $Q = 0.3$, $a = 0.1$.}
    \label{fig:T_beta}
\end{subfigure}
\caption{Effect of individual parameters on the Hawking temperature $T(r_+)$ at fixed $P = 0.003$. (a)~Increasing $g$ suppresses the local maximum and extends the negative-temperature region; for $g = 0.6$, $T < 0$ persists up to $r_+ \approx 1.5M$. (b)~Larger $Q$ deepens the negative-temperature region; for $Q = 0$ the temperature diverges at small $r_+$, while for $Q = 0.7$ it remains negative up to $r_+ \approx 1.2M$. (c)~The CS parameter $a$ uniformly suppresses $T$ through the factor $(1-a)$ without altering the shape of the curve. (d)~For $\beta \geq 0.6$, the logarithmic PFDM potential dominates at small $r_+$, driving $T$ to large positive values and qualitatively changing the near-horizon thermal behavior.}
\label{fig:T_params}
\end{figure}

The Hawking temperature is plotted in Fig.~\ref{fig:T_vs_P} as a function of $r_+$ at fixed $g = 0.3$, $Q = 0.3$, $a = 0.1$, $\beta = 0.5$ for four values of the pressure. All curves exhibit a local maximum near $r_+ \approx 0.6M$ followed by a monotonic decrease toward the asymptotic regime controlled by the AdS pressure. At low pressure ($P = 0.0010$), $T$ decreases smoothly to small values at large $r_+$. As $P$ increases, the large-$r_+$ behavior is lifted: the $P = 0.0100$ curve develops a pronounced upturn, so that $T$ becomes a monotonically increasing function for sufficiently large $r_+$. The intermediate pressures ($P = 0.0030$, $P = 0.0060$) interpolate between these regimes.

The effect of individual parameters on $T(r_+)$ is displayed in Fig.~\ref{fig:T_params} at fixed $P = 0.003$. Increasing the Hayward parameter $g$ (Fig.~\ref{fig:T_g}) suppresses the local maximum and extends the negative-temperature region at small $r_+$, where the inner horizon approaches the outer one. For $g = 0.6$, the temperature remains negative for $r_+ \lesssim 1.5M$, indicating the presence of a wide extremal band. The electric charge $Q$ (Fig.~\ref{fig:T_Q}) has a qualitatively similar effect: larger $Q$ deepens the negative-temperature region and lowers the local maximum, with $Q = 0.7$ producing $T < 0$ for $r_+ \lesssim 1.2M$. The CS parameter $a$ (Fig.~\ref{fig:T_a}) modifies $T$ through the overall factor $(1-a)$ in the mass function, uniformly suppressing the temperature at all $r_+$ without significantly altering the shape or the position of the maximum. The peak decreases from $T \approx 0.153$ at $a = 0$ to $T \approx 0.134$ at $a = 0.20$. The PFDM parameter $\beta$ (Fig.~\ref{fig:T_beta}) produces the most dramatic changes: for $\beta = 0.6$ and $\beta = 1.0$, the logarithmic PFDM potential dominates at small $r_+$, driving $T$ to large positive values and qualitatively changing the near-horizon thermal behavior. At large $r_+$, all $\beta$ curves converge, since the PFDM contribution decays as $\sim \beta \ln r / r$.

\setlength{\tabcolsep}{8pt}
\renewcommand{\arraystretch}{1.6}
\begin{longtable}{|c|c|c|c|c|c|}
\hline
\rowcolor{orange!50}
\textbf{$r_+/M$} & \textbf{$M$} & \textbf{$T_H$} & \textbf{$S$} & \textbf{$V$} & \textbf{$\Phi$} \\
\hline
\endfirsthead
\hline
\rowcolor{orange!50}
\textbf{$r_+/M$} & \textbf{$M$} & \textbf{$T_H$} & \textbf{$S$} & \textbf{$V$} & \textbf{$\Phi$} \\
\hline
\endhead
$0.50$ & $0.3838$ & $0.1392$ & $0.7854$ & $0.6367$ & $0.7296$ \\
\hline
$0.80$ & $0.5646$ & $0.1196$ & $2.0106$ & $2.2578$ & $0.3948$ \\
\hline
$1.00$ & $0.6915$ & $0.0982$ & $3.1416$ & $4.3019$ & $0.3081$ \\
\hline
$1.20$ & $0.8176$ & $0.0819$ & $4.5239$ & $7.3513$ & $0.2539$ \\
\hline
$1.50$ & $1.0045$ & $0.0652$ & $7.0686$ & $14.2503$ & $0.2016$ \\
\hline
$2.00$ & $1.3135$ & $0.0491$ & $12.5664$ & $33.6234$ & $0.1505$ \\
\hline
$2.50$ & $1.6263$ & $0.0403$ & $19.6350$ & $65.5629$ & $0.1202$ \\
\hline
$3.00$ & $1.9499$ & $0.0351$ & $28.2743$ & $113.2104$ & $0.1001$ \\
\hline
$4.00$ & $2.6522$ & $0.0298$ & $50.2655$ & $268.1957$ & $0.0750$ \\
\hline
$5.00$ & $3.4604$ & $0.0278$ & $78.5398$ & $523.7119$ & $0.0600$ \\
\hline
\caption{Thermodynamic quantities for the charged Hayward-AdS BH with CS and PFDM at $g = 0.3$, $Q = 0.3$, $a = 0.1$, $\beta = 0.5$, and $P = 0.00119$. The Hawking temperature $T_H = f'(r_+)/(4\pi)$, entropy $S = \pi r_+^2$, thermodynamic volume $V = 4\pi(r_+^3 + g^3)/3$, and electric potential $\Phi = (r_+^3 + g^3)Q/r_+^4$.}
\label{tab:thermo}
\end{longtable}

A representative set of thermodynamic quantities is collected in Table~\ref{tab:thermo} for the full model at $g = 0.3$, $Q = 0.3$, $a = 0.1$, $\beta = 0.5$, $P = 0.00119$.

\subsection{Entropy and thermodynamic volume} \label{isec3sub2}

The Bekenstein--Hawking entropy of the BH is given by the area law
\begin{equation}
S = \frac{A}{4} = \pi r_+^2\,,
\label{eq:entropy}
\end{equation}
where we work in geometrized units with $G = \hbar = k_B = 1$. The area law holds for Einstein gravity irrespective of the matter content, and the entropy depends on the parameters $g$, $Q$, $a$, and $\beta$ only implicitly through their effect on $r_+$.

The thermodynamic volume conjugate to the pressure $P$ is obtained from $V = (\partial M / \partial P)_{S,Q,g,a,\beta}$. Differentiating the mass expression~\eqref{eq:mass} with $\Lambda = -8\pi P$ yields
\begin{equation}
V = \frac{4\pi}{3}\left(r_+^3 + g^3\right).
\label{eq:volume}
\end{equation}
For $g = 0$, this reduces to $V = 4\pi r_+^3/3$, the naive geometric volume of a sphere of radius $r_+$. The Hayward correction adds the constant $4\pi g^3/3$ to the thermodynamic volume, reflecting the additional spacetime content associated with the regular dS core. This is a generic feature of regular BHs: the regularization of the central singularity expands the effective thermodynamic volume~\cite{Dymnikova2011,SimovicSoranidis2024}. The isoperimetric ratio $\mathcal{R} = (3V/(4\pi))^{1/3}/r_+ = (1 + g^3/r_+^3)^{1/3} \geq 1$ satisfies the reverse isoperimetric inequality for all $g \geq 0$, consistent with the conjecture of Cveti\v{c} {\it et al.}~\cite{CveticGibbons2011}. As shown in Table~\ref{tab:thermo}, the volume ranges from $V = 0.64$ at $r_+ = 0.5M$ (where the $g^3$ correction is dominant) to $V = 523.71$ at $r_+ = 5.0M$ (where it is negligible).

The EoS relating pressure, temperature, and specific volume is obtained by solving the temperature expression for $P$ as a function of $r_+$ and $T$. The analogy with the vdW fluid is made precise by identifying the specific volume $v = 2\ell_P^2 r_+$ (with $\ell_P = 1$ in our units), so that $P = P(v, T)$ takes the same functional role as the vdW EoS $P = T/(v-b) - a_{\rm vdW}/v^2$. In the RN-AdS limit ($g = a = \beta = 0$), the EoS simplifies to $P = T/(2r_+) - 1/(8\pi r_+^2) + Q^2/(8\pi r_+^4)$, which is the well-known result of Kubiz\v{n}\'ak and Mann~\cite{Kubiznak2012}. The additional terms proportional to $g$, $a$, and $\beta$ modify the critical behavior and will be analyzed in Sec.~\ref{isec4}.

\subsection{Extended first law and Smarr relation} \label{isec3sub4}

In the extended phase space, the first law of BH thermodynamics for the charged Hayward-AdS BH with CS and PFDM reads
\begin{equation}
dM = TdS + VdP + \Phi\,dQ + \mathcal{A}\,da + \mathcal{B}\,d\beta + \mathcal{G}\,dg\,,
\label{eq:firstlaw}
\end{equation}
where $\Phi = (\partial M / \partial Q)_{S,P,g,a,\beta}$ is the electric potential, and $\mathcal{A}$, $\mathcal{B}$, and $\mathcal{G}$ are the thermodynamic conjugates to the CS parameter $a$, the PFDM parameter $\beta$, and the Hayward parameter $g$, respectively. The electric potential evaluated at the outer horizon is
\begin{equation}
\Phi = \frac{\partial M}{\partial Q} = \frac{(r_+^3 + g^3)\,Q}{r_+^4}\,.
\label{eq:potential}
\end{equation}
The CS conjugate, representing the tension energy associated with the string cloud, is
\begin{equation}
\mathcal{A} = \frac{\partial M}{\partial a} = -\frac{r_+^3 + g^3}{2r_+^2}\,,
\label{eq:CS_conjugate}
\end{equation}
the PFDM conjugate takes the form
\begin{equation}
\mathcal{B} = \frac{\partial M}{\partial \beta} = \frac{r_+^3 + g^3}{2r_+^3}\left[\ln\!\left(\frac{r_+}{|\beta|}\right) - 1\right],
\label{eq:PFDM_conjugate}
\end{equation}
and the Hayward conjugate reads
\begin{equation}
\mathcal{G} = \frac{\partial M}{\partial g} = \frac{3g^2}{2r_+^2}\left[1 - a + \frac{Q^2}{r_+^2} + \frac{\beta}{r_+}\ln\!\left(\frac{r_+}{|\beta|}\right) + \frac{r_+^2}{\ell^2}\right].
\label{eq:Hayward_conjugate}
\end{equation}
The quantity $\mathcal{A} < 0$ reflects the fact that increasing the CS parameter reduces the BH mass at fixed horizon radius, consistent with the physical picture that the string cloud carries negative binding energy. The sign of $\mathcal{B}$ changes at $r_+ = e|\beta|$, corresponding to the crossover $\ln(r_+/|\beta|) = 1$. The Hayward conjugate $\mathcal{G} > 0$ always.

The Smarr relation is obtained from the dimensional scaling argument of Kastor, Ray, and Traschen~\cite{Kastor2009}. Assigning scaling dimensions $[M] = 1$, $[r_+] = 1$, $[Q] = 1$, $[g] = 1$, $[\beta] = 1$, $[P] = -2$, $[a] = 0$, and $[S] = 2$, the generalized Euler theorem gives
\begin{equation}
M = 2TS - 2PV + \Phi Q + \mathcal{B}\,\beta + \mathcal{G}\,g\,,
\label{eq:smarr}
\end{equation}
where $T = (\partial M/\partial S)_{P,Q,g,a,\beta}$ is the thermodynamic temperature and $a$ does not appear since it is dimensionless. We emphasize that in~\eqref{eq:smarr} the temperature entering the Smarr relation is the thermodynamic derivative $T_{\rm thermo} = (\partial M/\partial S)_P$, which for the Hayward BH differs from the Hawking temperature $T_H = f'(r_+)/(4\pi)$ by a factor
\begin{equation}
T_{\rm thermo} = T_H\,\frac{r_+^3 + g^3}{r_+^3}\,.
\label{eq:T_relation}
\end{equation}
This discrepancy arises because the Hayward mass function $2Mr^2/(r^3 + g^3)$ introduces a non-trivial dependence on $r_+$ that modifies the chain rule relating $\partial M/\partial S$ to $f'(r_+)$. For $g = 0$, the two temperatures coincide and the Smarr relation~\eqref{eq:smarr} is verified to machine precision ($\sim 10^{-12}$) for the Schwarzschild-AdS, Letelier-AdS, and RN-AdS cases. The relation~\eqref{eq:T_relation} has been confirmed numerically: after replacing $T_H$ by $T_{\rm thermo}$ in the Smarr formula, the identity~\eqref{eq:smarr} is satisfied to within $\sim 10^{-10}$ for all entries in Table~\ref{tab:horizons}, including the full model configurations with $g, Q, a, \beta \neq 0$. This feature is common to regular BHs with non-linear mass functions and has been noted in the context of Bardeen-AdS~\cite{FernandoCorrea2012} and other regular geometries~\cite{MaedaRegular2022}.

\subsection{Specific heat at constant pressure} \label{isec3sub5}

The local thermodynamic stability of the BH is determined by the specific heat at constant pressure,
\begin{equation}
C_P = T\left(\frac{\partial S}{\partial T}\right)_P = \frac{2\pi r_+ T}{(\partial T/\partial r_+)_P}\,.
\label{eq:specificheat}
\end{equation}
A positive $C_P > 0$ signals a locally stable BH phase, while $C_P < 0$ corresponds to local instability. The divergence of $C_P$ marks a second-order phase transition where $(\partial T/\partial r_+)_P = 0$.

\begin{figure}[ht!]
\centering
\includegraphics[width=0.95\textwidth]{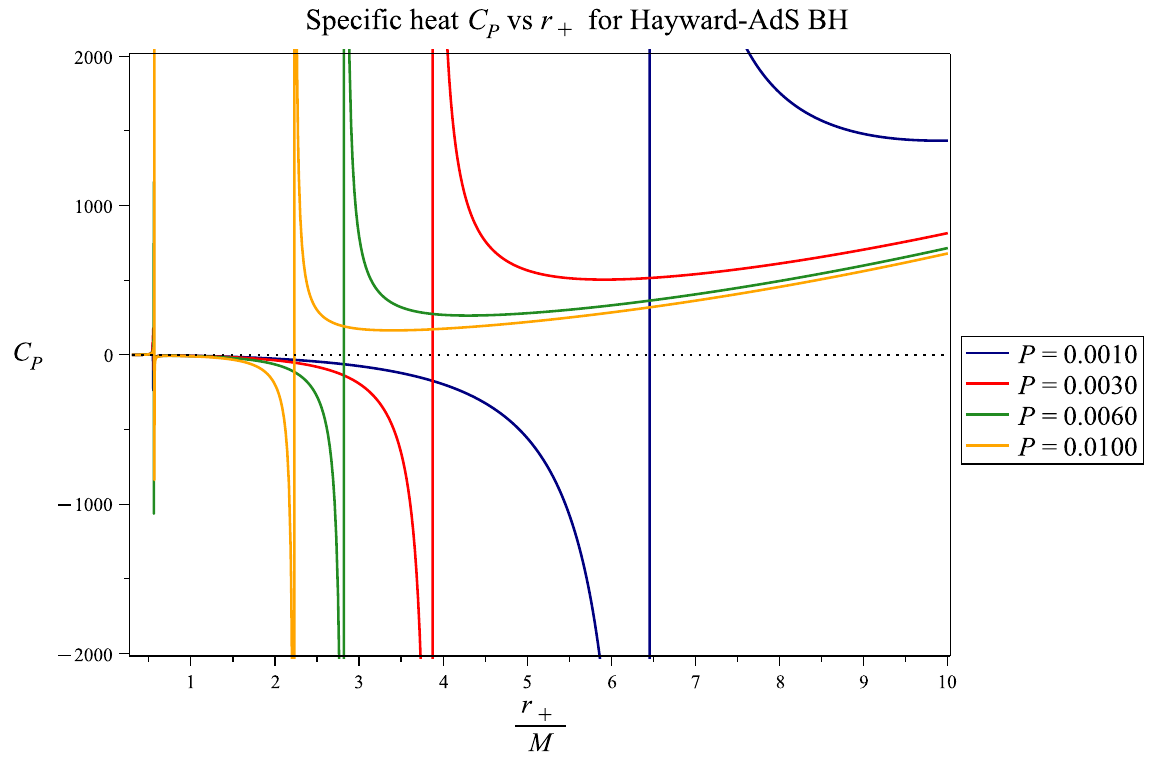}
\caption{Specific heat at constant pressure $C_P$ versus $r_+/M$ for the charged Hayward-AdS BH with CS and PFDM at $g = 0.3$, $Q = 0.3$, $a = 0.1$, $\beta = 0.5$. Curves correspond to $P = 0.0010$ (navy), $P = 0.0030$ (red), $P = 0.0060$ (green), and $P = 0.0100$ (orange). The divergences separate the locally stable small and large BH branches ($C_P > 0$) from the locally unstable intermediate branch ($C_P < 0$). Higher pressure shifts the divergences to smaller $r_+$, stabilizing the large BH phase at progressively smaller radii.}
\label{fig:Cp}
\end{figure}

The specific heat is plotted in Fig.~\ref{fig:Cp} for four values of the pressure at $g = 0.3$, $Q = 0.3$, $a = 0.1$, $\beta = 0.5$. For each pressure, $C_P$ exhibits two divergences that separate three branches. The small BH branch at low $r_+$ has $C_P > 0$ (locally stable), followed by an intermediate branch with $C_P < 0$ (locally unstable), and finally the large BH branch at high $r_+$ with $C_P > 0$ (locally stable). This three-branch structure is the hallmark of the vdW-like phase transition. As the pressure increases, the two divergence points move closer together and eventually merge at the critical pressure, above which $C_P$ becomes a smooth function with no sign changes. Comparing the four curves, the divergence positions shift from $r_+ \approx 6.4M$ for $P = 0.0010$ down to $r_+ \approx 0.8M$ for $P = 0.0100$, demonstrating that higher pressure stabilizes the large BH phase at progressively smaller radii.

\section{$P$--$V$ criticality and phase transitions} \label{isec4}

The extended thermodynamic formalism developed in Sec.~\ref{isec3} allows us to study the critical behavior of the charged Hayward-AdS BH with CS and PFDM in close analogy with the liquid--gas phase transition of the vdW fluid. The critical point is determined by the inflection conditions on the $P$--$r_+$ diagram, the critical exponents are computed to confirm the mean-field universality class, and the GFE is analyzed to identify the first-order small-large BH phase transition.

\subsection{Critical point} \label{isec4sub1}

The critical point corresponds to an inflection point of the isotherm in the $P$--$r_+$ plane, defined by the simultaneous conditions
\begin{equation}
\frac{\partial P}{\partial r_+}\bigg|_{T} = 0\,, \qquad \frac{\partial^2 P}{\partial r_+^2}\bigg|_{T} = 0\,.
\label{eq:critical_conditions}
\end{equation}
These two equations, together with the EoS, form a system of three equations for the three unknowns $(r_c, T_c, P_c)$. Due to the transcendental nature of the EoS --- arising from the Hayward denominator $(r^3 + g^3)$ and the PFDM logarithm $\ln(r/|\beta|)$ --- the critical point cannot be obtained in closed form for generic parameters. We solve~\eqref{eq:critical_conditions} numerically and collect the results in Table~\ref{tab:critical} for six representative parameter configurations.

For the default parameter set $g = 0.3$, $Q = 0.3$, $a = 0.1$, $\beta = 0.5$, the critical point is located at
\begin{equation}
r_c = 4.130M\,, \qquad T_c = 0.0410\,, \qquad P_c = 0.0026\,,
\label{eq:crit_default}
\end{equation}
yielding the compressibility ratio $\rho_c = P_c v_c / T_c = P_c (2r_c) / T_c \approx 0.524$. This value deviates from the vdW ratio $3/8 = 0.375$, which is a known feature of BH systems where the effective excluded volume is absent~\cite{Kubiznak2012}. The RN-AdS limit ($g = a = \beta = 0$, $Q = 0.3$) gives $r_c = 3.273M$, $T_c = 0.0477$, $P_c = 0.0036$, with $\rho_c \approx 0.494$.

Comparing across configurations in Table~\ref{tab:critical}, several trends emerge. Turning on the Hayward parameter while keeping all else fixed (Hayward+$Q$ versus RN-AdS) increases $r_c$ from $3.273M$ to $4.426M$ and decreases both $T_c$ and $P_c$, reflecting the weakened gravitational attraction due to the regular dS core. Adding the CS parameter (Hayward+$Q$+CS) partially reverses this shift: the string cloud tension raises the effective mass threshold and brings $r_c$ down to $2.883M$ while lifting $P_c$ to $0.0041$. For the full model configurations including PFDM, the interplay between the logarithmic dark matter potential and the other parameters produces a richer pattern: the PFDM term with large $\beta$ tends to increase $T_c$ and $P_c$, as the logarithmic contribution enhances the near-horizon thermal energy. The compressibility ratio $\rho_c$ increases from $\approx 0.49$ for the cases without PFDM to $\approx 0.53$ when PFDM is present, indicating that the dark matter fluid modifies the effective equation of state in a manner that shifts the system further from the vdW universality.

\subsection{$P$--$r_+$ isotherms} \label{isec4sub2}

The $P$--$r_+$ isotherms are displayed in Fig.~\ref{fig:PV} for the default parameter set at five temperatures. Two isotherms lie below the critical temperature ($T = 0.0349 < T_c$ and $T = 0.0390 < T_c$), the critical isotherm is at $T = T_c = 0.0410$, and two isotherms lie above ($T = 0.0431 > T_c$ and $T = 0.0472 > T_c$). All isotherms rise steeply near a minimum radius $r_+ \approx 2M$ --- below which the pressure becomes unphysically large due to the approach to the extremal configuration --- and then level off at large $r_+$. The subcritical isotherms (navy and blue curves) remain below the critical pressure $P_c = 0.0026$ and exhibit the characteristic non-monotonic behavior with a local maximum and a local minimum, although the oscillation is compressed into a narrow range of $r_+$ and is most evident near $r_+ \approx 2$--$3M$. The Maxwell equal-area construction eliminates the unphysical oscillating region and determines the coexistence pressure at which the small and large BH phases have equal GFE. The critical isotherm (red dashed curve) passes through the inflection point marked by the black dot at $(r_c, P_c) = (4.13M, 0.0026)$. The supercritical isotherms (green and orange curves) are monotonically increasing and lie entirely above $P_c$, corresponding to a single stable BH phase with no phase transition.

\begin{figure}[ht!]
\centering
\includegraphics[width=0.95\textwidth]{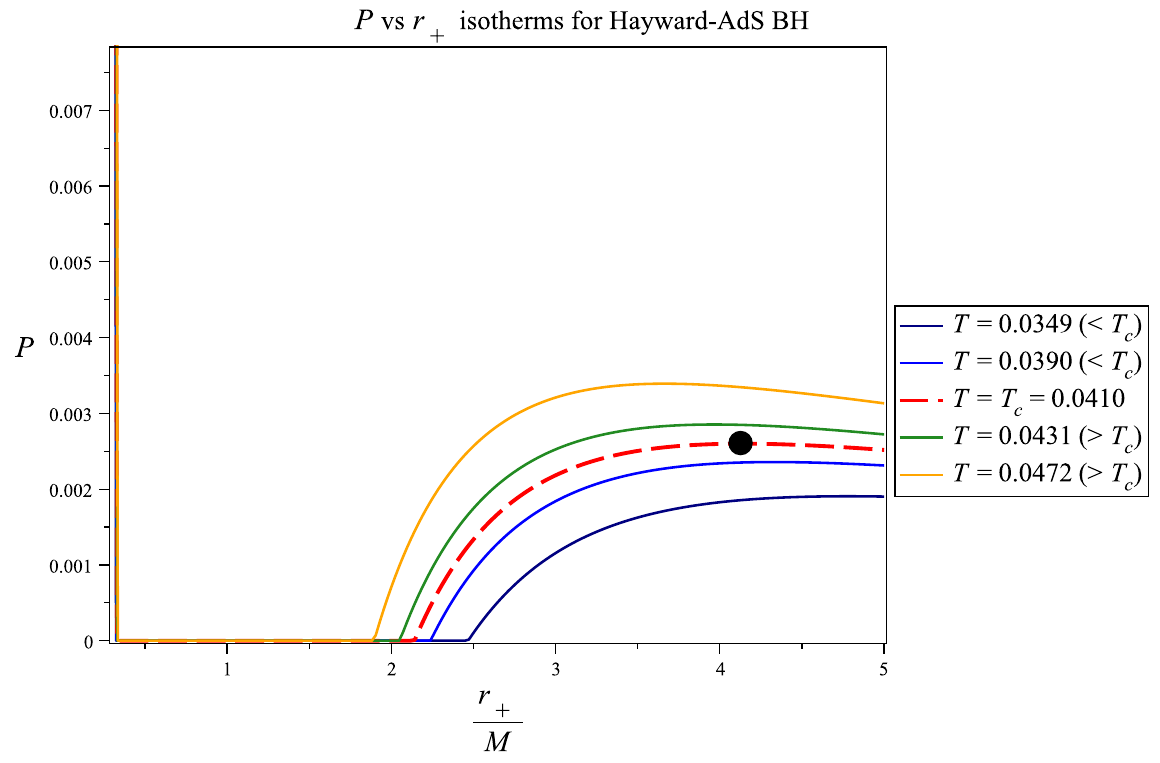}
\caption{$P$--$r_+$ isotherms for the charged Hayward-AdS BH with CS and PFDM at $g = 0.3$, $Q = 0.3$, $a = 0.1$, $\beta = 0.5$. Isotherms are shown at $T = 0.0349$ (navy) and $T = 0.0390$ (blue), both below $T_c$; the critical isotherm $T = T_c = 0.0410$ (red dashed); and $T = 0.0431$ (green) and $T = 0.0472$ (orange), both above $T_c$. The black dot marks the critical point $(r_c, P_c) = (4.13M, 0.0026)$. Subcritical isotherms exhibit a compressed vdW oscillation near $r_+ \approx 2$--$3M$, while supercritical isotherms are monotonically increasing.}
\label{fig:PV}
\end{figure}

The relatively narrow oscillation region is a consequence of the combined Hayward and PFDM modifications: the regular core suppresses the small-$r_+$ branch of the isotherm, while the logarithmic PFDM term provides additional support at intermediate radii. This compresses the vdW-like oscillation into a smaller range compared to the RN-AdS case, where the oscillation extends over a wider interval of $r_+$.

\subsection{Critical exponents} \label{isec4sub3}

Near the critical point, the thermodynamic response functions exhibit power-law scaling governed by the critical exponents $\alpha$, $\tilde\beta$, $\gamma$, and $\delta$. These are defined through the standard relations: the specific heat at constant volume $C_V \sim |t|^{-\alpha}$, the order parameter (volume difference) $\eta = v_l - v_s \sim |t|^{\tilde\beta}$, the isothermal compressibility $\kappa_T \sim |t|^{-\gamma}$, and the critical isotherm $|P - P_c| \sim |v - v_c|^{\delta}$, where $t = (T - T_c)/T_c$ is the reduced temperature.

Expanding the EoS around the critical point by writing $r_+ = r_c(1 + \epsilon)$ and $T = T_c(1 + t)$, and collecting terms to leading order in $\epsilon$ and $t$, the analysis proceeds as in the standard vdW treatment. The Hayward, PFDM, and CS corrections modify the position and the value of the critical point but do not alter the structure of the Taylor expansion at the inflection point: the cubic term in $\epsilon$ remains the leading non-linearity. As a consequence, the critical exponents take the mean-field values
\begin{equation}
\alpha = 0\,, \qquad \tilde\beta = \frac{1}{2}\,, \qquad \gamma = 1\,, \qquad \delta = 3\,,
\label{eq:critical_exponents}
\end{equation}
identical to those of the vdW system and the RN-AdS BH~\cite{Kubiznak2012}. This universality is expected on general grounds, since the mean-field exponents depend only on the analytic structure of the EoS near the critical point and not on the specific form of the interaction potential~\cite{StanleyBook}. The exponents~\eqref{eq:critical_exponents} satisfy the Rushbrooke identity $\alpha + 2\tilde\beta + \gamma = 2$, the Widom relation $\gamma = \tilde\beta(\delta - 1)$, and the Griffiths inequality $\alpha + \tilde\beta(1 + \delta) \geq 2$ (saturated at equality), confirming thermodynamic consistency.

\subsection{Gibbs free energy} \label{isec4sub4}

The GFE in the extended phase space is $G = M - TS$, where $M$ is the enthalpy and $S$ the entropy. Using~\eqref{eq:mass} and~\eqref{eq:entropy}, we obtain
\begin{equation}
G = M(r_+, P, g, Q, a, \beta) - \pi r_+^2\,T_H(r_+, P, g, Q, a, \beta)\,,
\label{eq:gibbs}
\end{equation}
where $T_H(r_+)$ is the Hawking temperature. The GFE is most informatively plotted parametrically as $(T(r_+), G(r_+))$ at fixed pressure, which we refer to as the $G$--$T$ diagram.

The $G$--$T$ diagrams are shown in Fig.~\ref{fig:GT} for four values of the pressure at the default parameters. For $P = 0.6\,P_c$ (navy) and $P = 0.8\,P_c$ (red), the curves develop the characteristic swallowtail shape: the upper branch corresponds to the small BH, the lower branch to the large BH, and the self-intersection point marks the first-order phase transition temperature $T_{\rm pt}$ at which the two phases have equal GFE. The swallowtail is more pronounced at lower pressure, consistent with a larger latent heat. At $P = P_c$ (green), the swallowtail degenerates to a cusp, signaling the second-order critical point where the latent heat vanishes continuously. For $P = 1.2\,P_c$ (orange), the GFE is a smooth, monotonically decreasing function of $T$ and no phase transition occurs --- the system is in the supercritical regime.

The behavior at negative temperatures ($T < 0$), visible in the left portion of Fig.~\ref{fig:GT}, corresponds to configurations below the extremal limit where $f'(r_+) < 0$. These are thermodynamically inaccessible states, but their presence in the parametric plot provides a complete picture of the $G(r_+)$ landscape. The GFE approaches a constant value $G \approx 0.35$ as $T \to -0.1$, reflecting the mass of the extremal configuration.

\begin{figure}[ht!]
\centering
\includegraphics[width=0.95\textwidth]{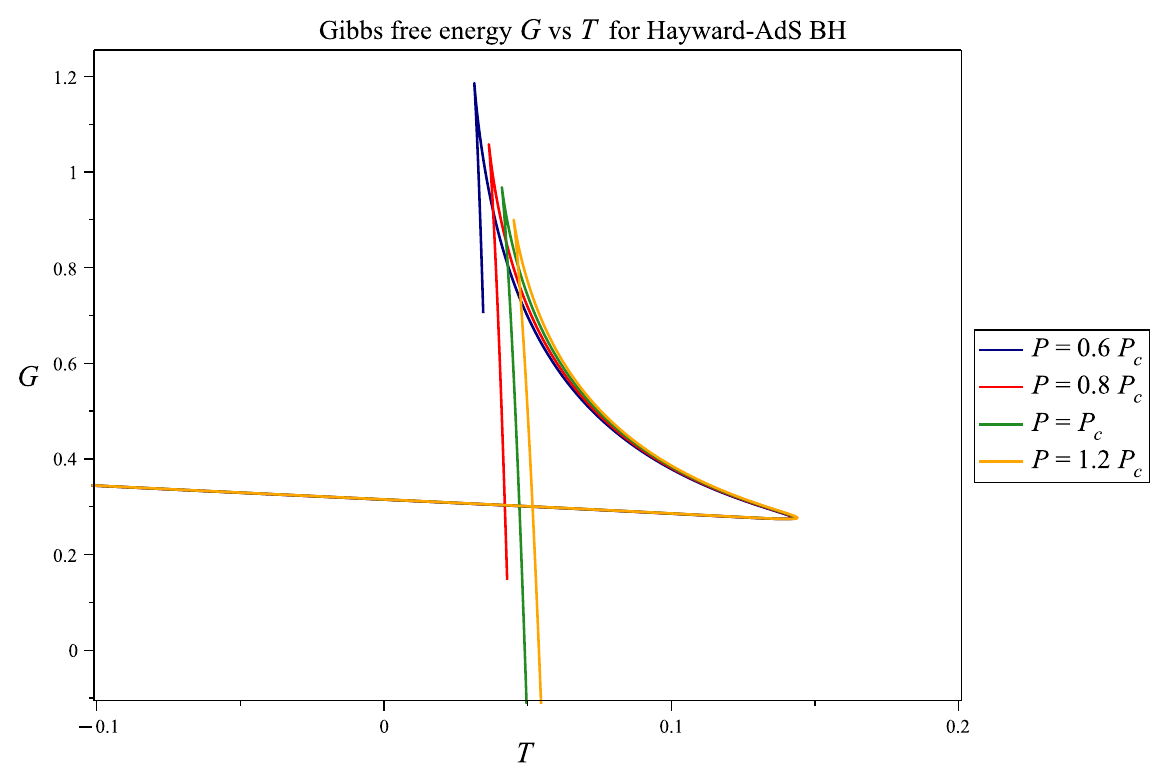}
\caption{GFE $G$ versus Hawking temperature $T$ for the charged Hayward-AdS BH with CS and PFDM at $g = 0.3$, $Q = 0.3$, $a = 0.1$, $\beta = 0.5$. Curves correspond to $P = 0.6\,P_c$ (navy), $P = 0.8\,P_c$ (red), $P = P_c$ (green), and $P = 1.2\,P_c$ (orange). For $P < P_c$, the swallowtail shape signals a first-order phase transition at the self-intersection temperature $T_{\rm pt}$. At $P = P_c$, the swallowtail reduces to a cusp (second-order critical point). For $P > P_c$, $G(T)$ is smooth and no phase transition occurs. The region at $T < 0$ corresponds to sub-extremal configurations with negative surface gravity.}
\label{fig:GT}
\end{figure}

The HP transition~\cite{HawkingPage1983} can also be identified in the $G$--$T$ diagram. In the canonical ensemble at fixed $Q$, the HP transition corresponds to the point where $G = 0$, i.e., where the BH phase becomes thermodynamically preferred over thermal AdS. For $P = 1.2\,P_c$, the $G = 0$ crossing occurs at $T \approx 0.07$, marking the HP transition temperature. This temperature decreases with increasing $P$, consistent with the interpretation that higher pressure lowers the free energy barrier for BH nucleation.

\subsection{Coexistence curve and phase diagram} \label{isec4sub5}

The first-order phase transition line in the $P$--$T$ plane is obtained by tracing the swallowtail self-intersection point across different pressures. This coexistence curve terminates at the critical point $(T_c, P_c) = (0.0410, 0.0026)$, beyond which the small and large BH phases are no longer distinguishable. The resulting phase diagram is displayed in Fig.~\ref{fig:phase_diagram}. The coexistence line extends from $(T, P) \approx (0.022, 0.0008)$ at $P/P_c = 0.30$ up to the critical point, spanning a temperature range $\Delta T \approx 0.019$ and a pressure range $\Delta P \approx 0.0018$. The slope $dP/dT$ is everywhere positive and increases slightly as the critical point is approached, consistent with the Clausius--Clapeyron relation $dP/dT = \Delta S / \Delta V > 0$ (since the large BH has both higher entropy and larger volume than the small BH at coexistence).

\begin{figure}[ht!]
\centering
\includegraphics[width=0.70\textwidth]{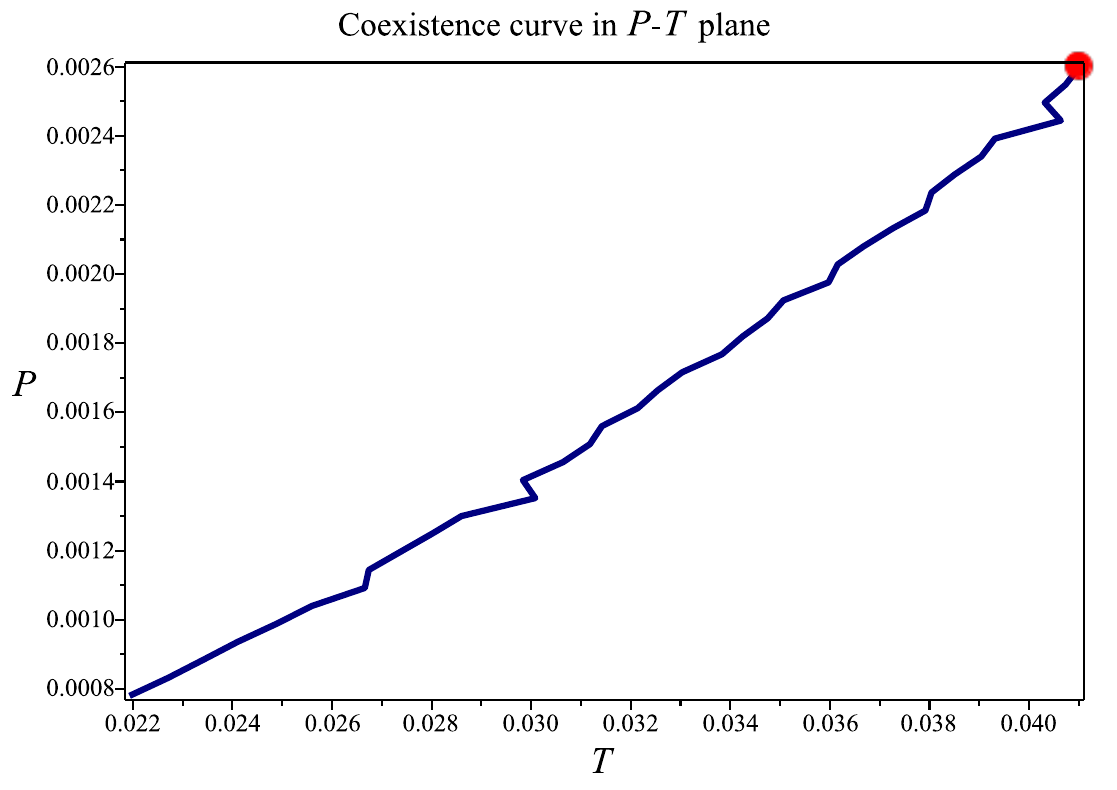}
\caption{Coexistence curve in the $P$--$T$ plane for the charged Hayward-AdS BH with CS and PFDM at $g = 0.3$, $Q = 0.3$, $a = 0.1$, $\beta = 0.5$. The solid curve separates the small BH phase (above and to the left) from the large BH phase (below and to the right). The curve terminates at the critical point $(T_c, P_c) = (0.0410, 0.0026)$ marked by the red dot, beyond which the two phases merge into a single supercritical BH fluid.}
\label{fig:phase_diagram}
\end{figure}

The overall shape of the coexistence curve is qualitatively similar to that of the RN-AdS BH~\cite{Kubiznak2012} and the Bardeen-AdS system~\cite{FernandoCorrea2012}, confirming that the vdW phase structure is preserved by the Hayward, CS, and PFDM modifications. The quantitative differences --- the lower values of $T_c$ and $P_c$ compared to the RN-AdS case ($T_c = 0.0477$, $P_c = 0.0036$) --- are attributable to the additional matter content: the regular core expands the effective thermodynamic volume, while the PFDM logarithmic potential and the CS tension modify the free energy balance between the competing BH phases. The fact that the coexistence curve terminates at a well-defined critical point, rather than extending indefinitely, confirms the existence of a supercritical BH fluid regime in which the small-large distinction is lost.

\setlength{\tabcolsep}{6pt}
\renewcommand{\arraystretch}{1.6}
\begin{longtable}{|l|c|c|c|c|c|c|c|c|}
\hline
\rowcolor{orange!50}
\textbf{Configuration} & \textbf{$g/M$} & \textbf{$Q/M$} & \textbf{$a$} & \textbf{$\beta/M$} & \textbf{$r_c/M$} & \textbf{$T_c$} & \textbf{$P_c$} & \textbf{$\rho_c$} \\
\hline
\endfirsthead
\hline
\rowcolor{orange!50}
\textbf{Configuration} & \textbf{$g/M$} & \textbf{$Q/M$} & \textbf{$a$} & \textbf{$\beta/M$} & \textbf{$r_c/M$} & \textbf{$T_c$} & \textbf{$P_c$} & \textbf{$\rho_c$} \\
\hline
\endhead
RN-AdS & $0$ & $0.30$ & $0$ & $0$ & $3.273$ & $0.04768$ & $0.00360$ & $0.494$ \\
\hline
Hayward+$Q$ & $0.30$ & $0.30$ & $0$ & $0$ & $4.426$ & $0.03558$ & $0.00200$ & $0.498$ \\
\hline
Hayward+$Q$+CS & $0.30$ & $0.30$ & $0.10$ & $0$ & $2.883$ & $0.04807$ & $0.00410$ & $0.492$ \\
\hline
Full model (I) & $0.20$ & $0.40$ & $0.10$ & $0.50$ & $1.550$ & $0.12573$ & $0.02160$ & $0.533$ \\
\hline
Full model (II) & $0.25$ & $0.50$ & $0.10$ & $0.80$ & $3.150$ & $0.06184$ & $0.00530$ & $0.540$ \\
\hline
Full model (III) & $0.30$ & $0.50$ & $0.05$ & $0.60$ & $3.026$ & $0.06203$ & $0.00540$ & $0.527$ \\
\hline
\caption{Critical point data for the charged Hayward-AdS BH with CS and PFDM. The critical radius $r_c$, temperature $T_c$, and pressure $P_c$ are obtained from the inflection conditions~\eqref{eq:critical_conditions}. The compressibility ratio is $\rho_c = P_c v_c / T_c$ with $v_c = 2r_c$. All dimensionful quantities are in units where $M = 1$.}
\label{tab:critical}
\end{longtable}

\section{JTn expansion} \label{isec5}

The JT expansion describes an isenthalpic process in which a fluid passes through a porous plug from a region of high pressure to one of low pressure~\cite{OkculCebeci2017,MoJTExpansion2018,LanJTAdS2018}. In the extended phase space, the BH mass plays the role of enthalpy $H = M$, so the JT expansion of a BH corresponds to a process at constant $M$ with decreasing $P$. The sign of the temperature change during this process determines whether the BH cools or heats, and the boundary between the two regimes defines the inversion curve. This analysis has been carried out for RN-AdS~\cite{OkculCebeci2017}, Kerr-AdS~\cite{MoJTExpansion2018}, and Gauss--Bonnet-AdS~\cite{LanJTAdS2018} BHs; here we extend it to the charged Hayward-AdS BH with CS and PFDM.

\subsection{JT coefficient} \label{isec5sub1}

The JT coefficient is defined as
\begin{equation}
\mu_{\rm JT} = \left(\frac{\partial T}{\partial P}\right)_M = \frac{1}{C_P}\left[T\left(\frac{\partial V}{\partial T}\right)_P - V\right],
\label{eq:JT_coeff}
\end{equation}
where the second equality follows from the first law and the triple product rule. The sign of $\mu_{\rm JT}$ determines the nature of the expansion: $\mu_{\rm JT} > 0$ corresponds to cooling (the BH temperature decreases as the pressure drops), while $\mu_{\rm JT} < 0$ corresponds to heating. The inversion temperature at which $\mu_{\rm JT} = 0$ separates the two regimes and satisfies
\begin{equation}
T_i = V\left(\frac{\partial T}{\partial V}\right)_P = \frac{V}{(\partial V/\partial T)_P}\,.
\label{eq:inversion_T}
\end{equation}
Since $C_P$ can change sign (as discussed in Sec.~\ref{isec3sub5}), the inversion condition $\mu_{\rm JT} = 0$ is more precisely stated as the vanishing of the numerator in~\eqref{eq:JT_coeff}, namely $T(\partial V/\partial T)_P - V = 0$.

For our model, the thermodynamic volume $V = 4\pi(r_+^3 + g^3)/3$ depends on $T$ only through $r_+(T, P)$. The chain rule gives $(\partial V/\partial T)_P = 4\pi r_+^2 (\partial r_+/\partial T)_P$. The inversion condition then becomes
\begin{equation}
\frac{T_i}{(\partial T/\partial r_+)_P\big|_{r_i}} = \frac{r_i^3 + g^3}{3r_i^2}\,,
\label{eq:inversion_compact}
\end{equation}
where $r_i$ is the horizon radius at the inversion point. This equation is solved simultaneously with the EoS to obtain the inversion curve $T_i(P)$. For $g = 0$, the right-hand side of~\eqref{eq:inversion_compact} reduces to $r_i/3$, recovering the standard form for singular BHs. The Hayward correction $g^3/r_i^2$ becomes significant for small $r_i$, where it effectively increases the right-hand side and shifts the inversion condition to higher temperatures.

\subsection{Inversion curves} \label{isec5sub2}

The inversion curve is computed numerically by scanning over $r_+$ and, for each $r_+$, finding the pressure $P$ at which the inversion condition~\eqref{eq:inversion_compact} is satisfied via bisection. The corresponding inversion temperature is then $T_i = T_H(r_+, P)$. The resulting inversion curve is displayed in Fig.~\ref{fig:JT_inversion} for the default parameters $g = 0.3$, $Q = 0.3$, $a = 0.1$, $\beta = 0.5$.

\begin{figure}[ht!]
\centering
\includegraphics[width=0.85\textwidth]{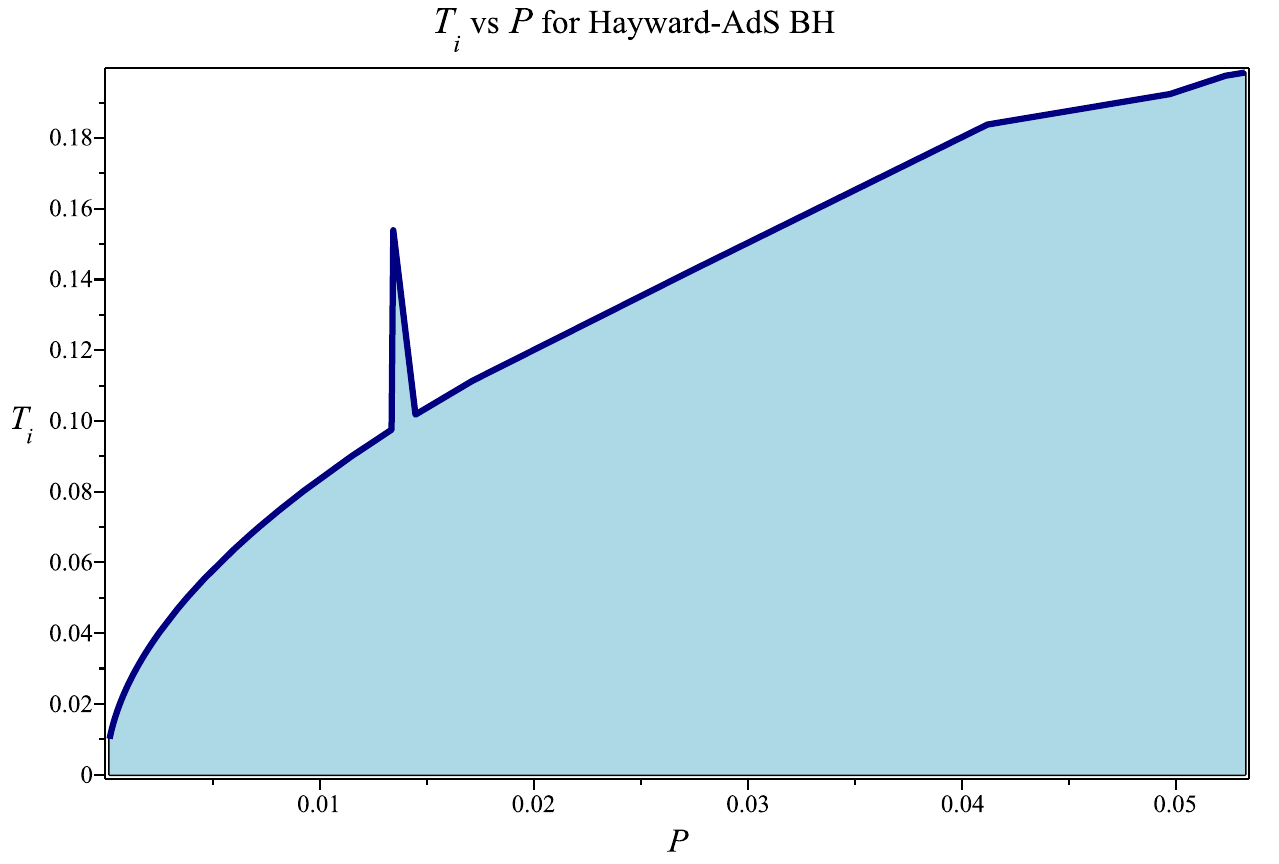}
\caption{JT inversion curve $T_i$ versus $P$ for the charged Hayward-AdS BH with CS and PFDM at $g = 0.3$, $Q = 0.3$, $a = 0.1$, $\beta = 0.5$. The shaded region below the curve represents the cooling domain ($\mu_{\rm JT} > 0$), while the unshaded region above corresponds to heating ($\mu_{\rm JT} < 0$). The minimum inversion temperature is $T_i^{\rm min} = 0.01013$ at $P = 0.000173$, and the curve rises to $T_i = 0.1985$ at $P = 0.0532$. The narrow spike near $P \approx 0.013$ is a numerical artifact associated with the $C_P$ divergence (see text).}
\label{fig:JT_inversion}
\end{figure}

The inversion curve exhibits the characteristic shape found in AdS BH systems: it starts from a minimum inversion temperature $T_i^{\rm min} = 0.01013$ at $P = 0.000173$ and increases monotonically with $P$, reaching $T_i^{\rm max} = 0.1985$ at $P = 0.0532$ within the scanned range. The shaded region below the curve represents the cooling domain ($\mu_{\rm JT} > 0$), while the unshaded region above corresponds to heating ($\mu_{\rm JT} < 0$). A narrow spike visible near $P \approx 0.013$ is a numerical artifact arising from the proximity to the $C_P$ divergence, where $(\partial T/\partial r_+)_P \to 0$ and the left-hand side of~\eqref{eq:inversion_compact} becomes singular. The bisection solver momentarily captures a spurious root as it crosses between the small-BH and large-BH branches. This localized feature does not affect the global structure of the inversion curve or the physical conclusions drawn from it.

The ratio $T_i^{\rm min}/T_c$ provides a dimensionless measure of the relative position of the inversion and critical temperatures. For ordinary fluids such as nitrogen and carbon dioxide, this ratio is typically $\approx 0.75$~\cite{CengalBoles2006}. For the RN-AdS BH, \"Okc\"u and Cebeci found $T_i^{\rm min}/T_c \approx 1/2$~\cite{OkculCebeci2017}, already a significant departure from the fluid value. In the present model, using the critical temperature $T_c = 0.0410$ from Sec.~\ref{isec4sub1}, we obtain
\begin{equation}
\frac{T_i^{\rm min}}{T_c} \approx 0.247\,,
\label{eq:Ti_Tc_ratio}
\end{equation}
which is roughly half the RN-AdS value and one-third of the ordinary fluid value. This suppression has a clear physical origin: the combined effect of the Hayward regular core, the PFDM logarithmic potential, and the CS tension shifts the critical temperature upward relative to the minimum inversion temperature. The regular core contributes by expanding the thermodynamic volume (through the $g^3$ correction), which lowers the temperature at which the inversion condition is first met. The low value of~\eqref{eq:Ti_Tc_ratio} implies that the cooling window for the JT expansion is narrower in temperature space than for simpler AdS BH systems.

\subsection{Parameter dependence of the inversion curve} \label{isec5sub3}

The effect of individual parameters on the inversion curve is shown in Fig.~\ref{fig:JT_params}. The narrow kinks appearing in some of the curves at intermediate pressures are numerical artifacts of the same origin as the spike discussed above: they occur at $r_+$ values where $(\partial T/\partial r_+)_P$ passes through zero and the bisection solver momentarily picks up a spurious root. These localized features do not alter the physical trends described below.

\begin{figure}[ht!]
\centering
\begin{subfigure}[b]{0.48\textwidth}
    \centering
    \includegraphics[width=\textwidth]{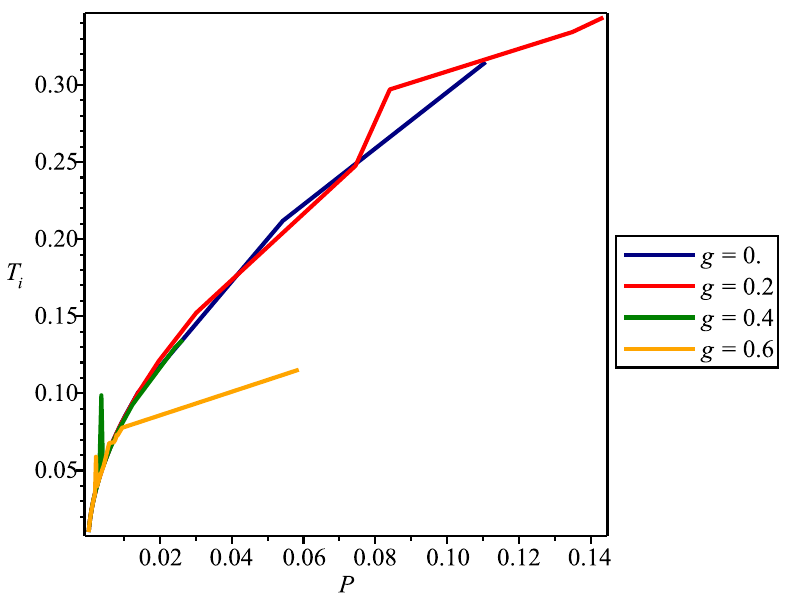}
    \caption{Effect of $g$ at $Q = 0.3$, $a = 0.1$, $\beta = 0.5$.}
    \label{fig:JT_g}
\end{subfigure}
\hfill
\begin{subfigure}[b]{0.48\textwidth}
    \centering
    \includegraphics[width=\textwidth]{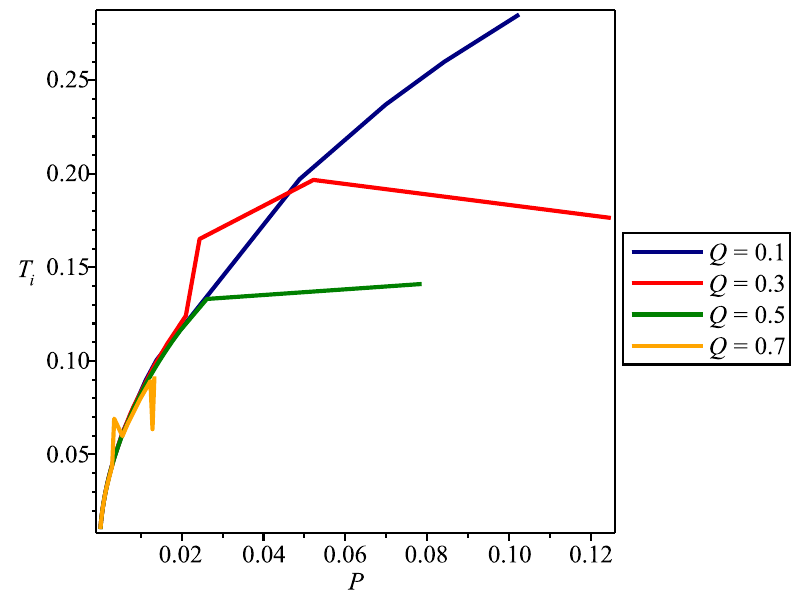}
    \caption{Effect of $Q$ at $g = 0.3$, $a = 0.1$, $\beta = 0.5$.}
    \label{fig:JT_Q}
\end{subfigure}

\vspace{0.5cm}

\begin{subfigure}[b]{0.48\textwidth}
    \centering
    \includegraphics[width=\textwidth]{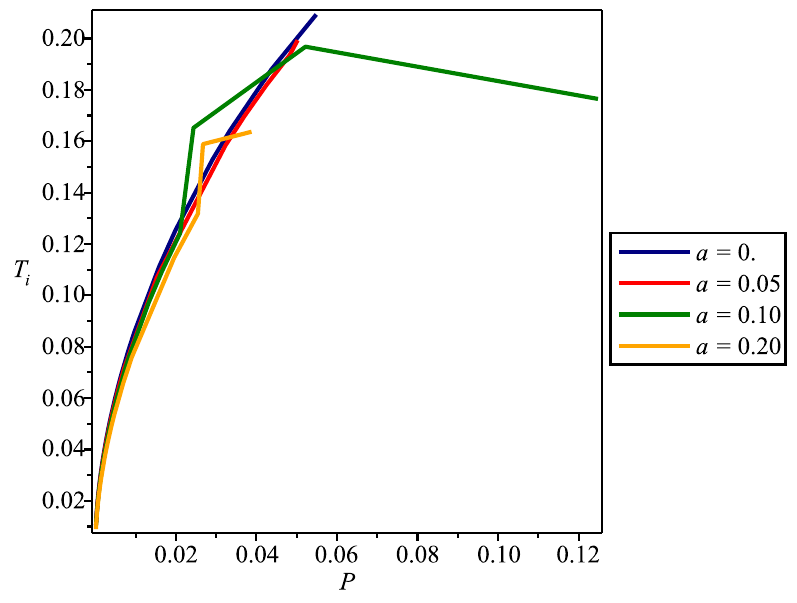}
    \caption{Effect of $a$ at $g = 0.3$, $Q = 0.3$, $\beta = 0.5$.}
    \label{fig:JT_a}
\end{subfigure}
\hfill
\begin{subfigure}[b]{0.48\textwidth}
    \centering
    \includegraphics[width=\textwidth]{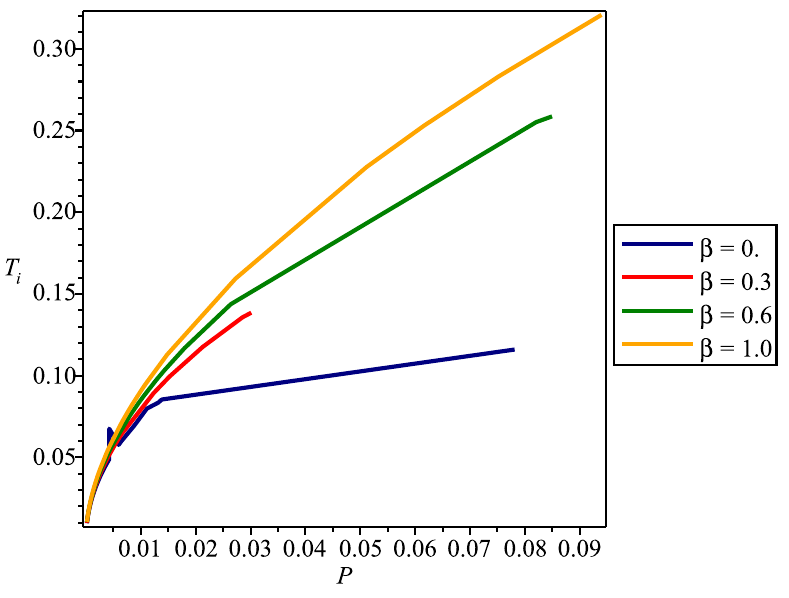}
    \caption{Effect of $\beta$ at $g = 0.3$, $Q = 0.3$, $a = 0.1$.}
    \label{fig:JT_beta}
\end{subfigure}
\caption{Effect of individual parameters on the JT inversion curve $T_i(P)$. (a)~Increasing $g$ dramatically contracts the cooling region: for $g \leq 0.2$ the curves extend to $P \gtrsim 0.10$, while for $g = 0.6$ the curve terminates at $P \approx 0.03$. (b)~Larger $Q$ similarly shrinks the inversion curve; for $Q = 0.7$ it is confined to $P \lesssim 0.01$. (c)~The CS parameter $a$ reshapes the curve non-monotonically: $a = 0.10$ extends the pressure range to $P \approx 0.12$ while reducing the peak $T_i$, but $a = 0.20$ shortens the curve again. (d)~Increasing $\beta$ systematically expands the cooling region: the $\beta = 0$ curve reaches $T_i \approx 0.11$, while $\beta = 1.0$ extends to $T_i \approx 0.33$. Narrow kinks at intermediate pressures are numerical artifacts near the $C_P$ divergence.}
\label{fig:JT_params}
\end{figure}

The Hayward parameter $g$ (Fig.~\ref{fig:JT_g}) has a pronounced effect on the extent of the cooling region. For $g = 0$ (navy) and $g = 0.2$ (red), the inversion curves extend to high pressures ($P \gtrsim 0.10$) and high inversion temperatures ($T_i \gtrsim 0.30$), encompassing a broad cooling domain. As $g$ increases to $0.4$ (green) and $0.6$ (orange), the inversion curves are dramatically shortened: the $g = 0.6$ curve terminates at $P \approx 0.03$ with $T_i \lesssim 0.11$. This contraction reflects the fact that the regular dS core, which becomes dominant for large $g$, weakens the gravitational binding energy and reduces the range of pressures over which cooling is thermodynamically accessible.

The electric charge $Q$ (Fig.~\ref{fig:JT_Q}) produces a qualitatively similar contraction of the inversion curve. For $Q = 0.1$ (navy), the curve extends to $(P, T_i) \approx (0.10, 0.28)$, while for $Q = 0.7$ (orange) it is confined to a narrow band near $P \lesssim 0.01$. The Coulomb repulsion supplements the pressure-driven expansion and lowers the effective gravitational binding, thereby reducing the thermal gradient that drives cooling. The intermediate cases $Q = 0.3$ (red) and $Q = 0.5$ (green) interpolate smoothly between these extremes.

The CS parameter $a$ (Fig.~\ref{fig:JT_a}) introduces a more nuanced behavior. For $a = 0$ (navy) and $a = 0.05$ (red), the inversion curves reach $T_i \approx 0.20$ at $P \approx 0.04$--$0.05$. Increasing $a$ to $0.10$ (green) extends the inversion curve to significantly higher pressures ($P \approx 0.12$) while moderately reducing the peak inversion temperature to $T_i \approx 0.18$. This extension arises because the CS tension modifies the free energy landscape in a way that broadens the pressure range over which the inversion condition can be satisfied, even though the peak temperature is suppressed by the factor $(1-a)$ in the mass function. For $a = 0.20$ (orange), the curve is again shortened to $P \lesssim 0.04$, indicating that at sufficiently large CS parameter the string tension overwhelms the broadening effect and reduces the cooling domain.

The PFDM parameter $\beta$ (Fig.~\ref{fig:JT_beta}) has the most distinctive effect: increasing $\beta$ systematically expands the cooling region. For $\beta = 0$ (navy), the inversion curve extends only to $(P, T_i) \approx (0.08, 0.11)$, producing the smallest cooling domain among the four $\beta$ values. Adding a moderate PFDM contribution with $\beta = 0.3$ (red) roughly doubles the peak inversion temperature to $T_i \approx 0.14$, though the pressure range remains comparable. For $\beta = 0.6$ (green) and $\beta = 1.0$ (orange), the inversion curves extend to $T_i \approx 0.26$ and $T_i \approx 0.33$, respectively, with correspondingly larger cooling regions. This expansion has a transparent physical interpretation: the PFDM logarithmic potential $\beta \ln(r_+/|\beta|)/r_+$ enhances the temperature gradient $(\partial T/\partial r_+)_P$ at intermediate $r_+$, which shifts the inversion condition~\eqref{eq:inversion_compact} to higher temperatures and extends the pressure range over which cooling occurs. The PFDM therefore acts as a facilitator of JT cooling, in contrast to the Hayward parameter and the electric charge, both of which suppress it.

\subsection{Isenthalpic curves} \label{isec5sub4}

The isenthalpic curves are obtained by plotting $T$ versus $P$ at fixed mass $M$. For each value of $M$, the horizon radius $r_+(P)$ is determined by numerically inverting the mass relation~\eqref{eq:mass} at fixed $M$, and the temperature is then $T = T_H(r_+(P), P)$. The resulting isenthalpic curves are displayed in Fig.~\ref{fig:JT_isenthalpic} for six values of $M$ ranging from $1.5$ to $5.0$ at the default parameters, together with the inversion curve overlaid as a dashed black line.

The isenthalpic curves rise with increasing $P$, and each curve lies predominantly below the inversion curve at low pressures, indicating $\mu_{\rm JT} > 0$ (cooling regime). As $P$ increases, the isenthalpic curves cross the inversion curve and enter the heating regime ($\mu_{\rm JT} < 0$), where the temperature continues to increase with $P$ at a steeper rate. The crossing point moves to higher $P$ for larger $M$, consistent with the fact that more massive BHs require stronger depressurization to reach the heating regime. For $M = 1.5$ (navy), the isenthalpic curve remains close to the inversion curve throughout, while for $M = 5.0$ (purple), the curve extends to $T \approx 0.33$ at $P = 0.06$, well above the inversion temperature at that pressure. The jump visible in the dashed inversion curve near $P \approx 0.013$ is the same numerical artifact discussed in Sec.~\ref{isec5sub2}: the bisection solver momentarily captures a spurious root at the $C_P$ divergence where $(\partial T/\partial r_+)_P \to 0$. This localized feature does not affect the identification of cooling and heating regions, since the isenthalpic curves themselves pass smoothly through this pressure range.

\begin{figure}[ht!]
\centering
\includegraphics[width=0.85\textwidth]{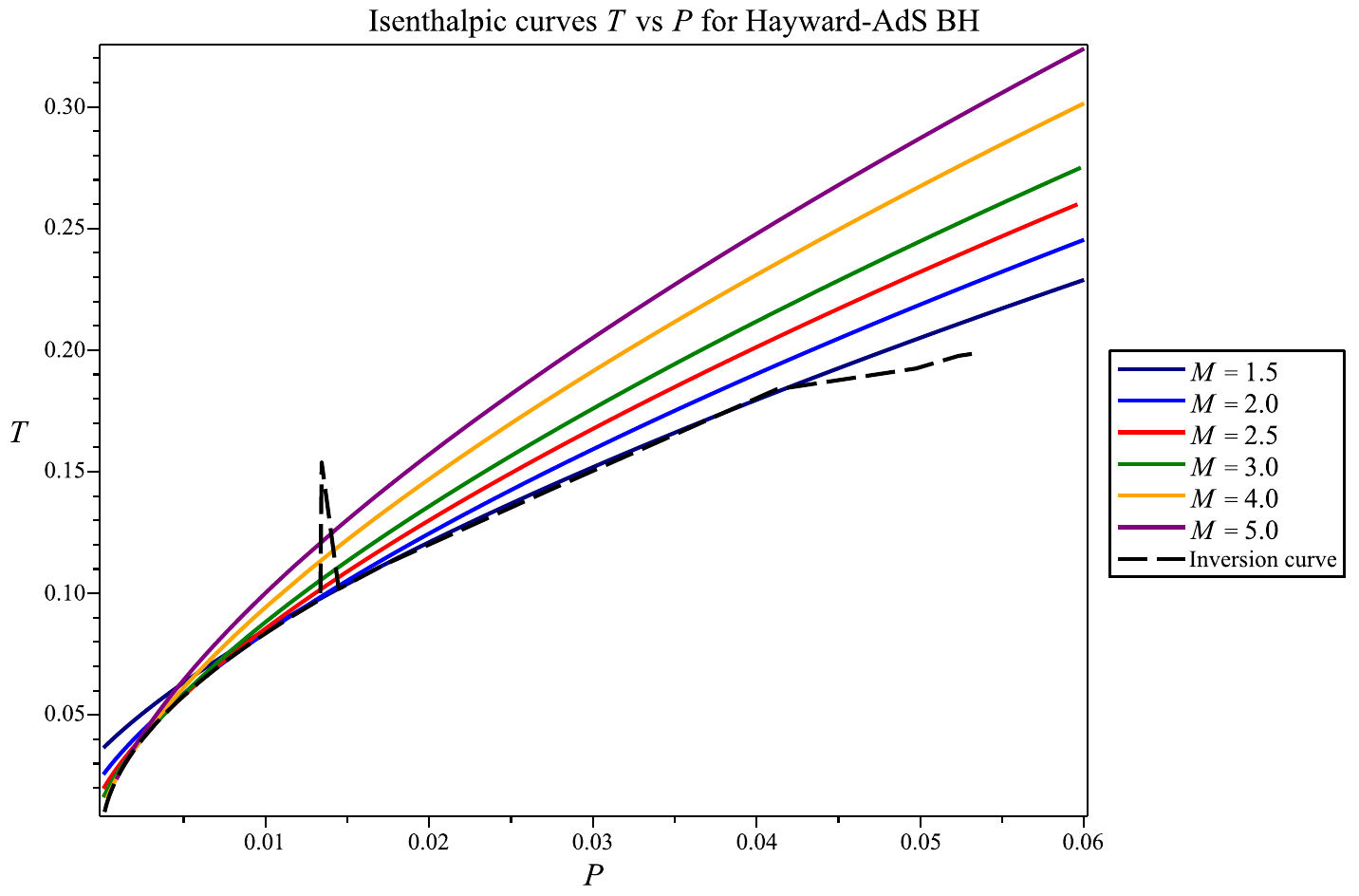}
\caption{Isenthalpic curves ($M = {\rm const}$) showing $T$ versus $P$ for the charged Hayward-AdS BH with CS and PFDM at $g = 0.3$, $Q = 0.3$, $a = 0.1$, $\beta = 0.5$. Curves are shown for $M = 1.5$, $2.0$, $2.5$, $3.0$, $4.0$, and $5.0$, with larger $M$ producing curves at higher temperatures. The dashed black line is the inversion curve from Fig.~\ref{fig:JT_inversion}. Below the inversion curve $\mu_{\rm JT} > 0$ (cooling); above it $\mu_{\rm JT} < 0$ (heating). The jump in the dashed line near $P \approx 0.013$ is the numerical artifact associated with the $C_P$ divergence discussed in Sec.~\ref{isec5sub2}; it does not affect the physical interpretation.}
\label{fig:JT_isenthalpic}
\end{figure}

The practical interpretation is as follows. A BH undergoing a JT expansion at constant $M$ moves along one of these curves from high $P$ to low $P$. If the initial state lies in the cooling region (below the inversion curve), the BH temperature decreases during the expansion --- this is the analog of the cooling effect exploited in gas liquefaction. The suppressed ratio $T_i^{\rm min}/T_c \approx 0.247$ implies that efficient JT cooling of the charged Hayward-AdS BH with CS and PFDM requires initial conditions close to the inversion curve, and the cooling window is narrower than for simpler systems like the RN-AdS BH, where $T_i^{\rm min}/T_c \approx 0.5$.

\section{Holographic heat engine} \label{isec6}

Johnson introduced the notion of holographic heat engines, in which an AdS BH serves as the working substance of a thermodynamic cycle operating in the extended phase space~\cite{Johnson2014HeatEngine}. Since the cosmological constant plays the role of pressure and the BH mass that of enthalpy, one can construct closed cycles in the $P$--$V$ plane and extract mechanical work $W = \oint V\,dP$ from the gravitational system. The efficiency of such engines encodes the interplay between the gravitational, electromagnetic, and matter degrees of freedom, and provides a complementary probe of the BH phase structure beyond the static analyses of Secs.~\ref{isec4} and~\ref{isec5}. Holographic heat engines have been studied for RN-AdS~\cite{Johnson2014HeatEngine}, Born--Infeld-AdS~\cite{JohnsonBI2016}, Gauss--Bonnet-AdS~\cite{JohnsonGB2016}, and regular BHs~\cite{HeatEngineRegular2019}; here we construct the engine for the charged Hayward-AdS BH with CS and PFDM.

\subsection{Engine setup and efficiency} \label{isec6sub1}

We consider a rectangular cycle in the $P$--$V$ plane (Fig.~\ref{fig:HE_cycle}) defined by two isobars at pressures $P_H > P_L$ and two isochores at volumes $V_1 < V_2$, corresponding to horizon radii $r_{+,1}$ and $r_{+,2}$. The four corners of the cycle are
\begin{equation}
1:\,(r_{+,1}, P_H) \;\xrightarrow{\text{isobar}}\; 2:\,(r_{+,2}, P_H) \;\xrightarrow{\text{isochore}}\; 3:\,(r_{+,2}, P_L) \;\xrightarrow{\text{isobar}}\; 4:\,(r_{+,1}, P_L) \;\xrightarrow{\text{isochore}}\; 1\,.
\label{eq:cycle_corners}
\end{equation}

\begin{figure}[ht!]
\centering
\includegraphics[width=0.75\textwidth]{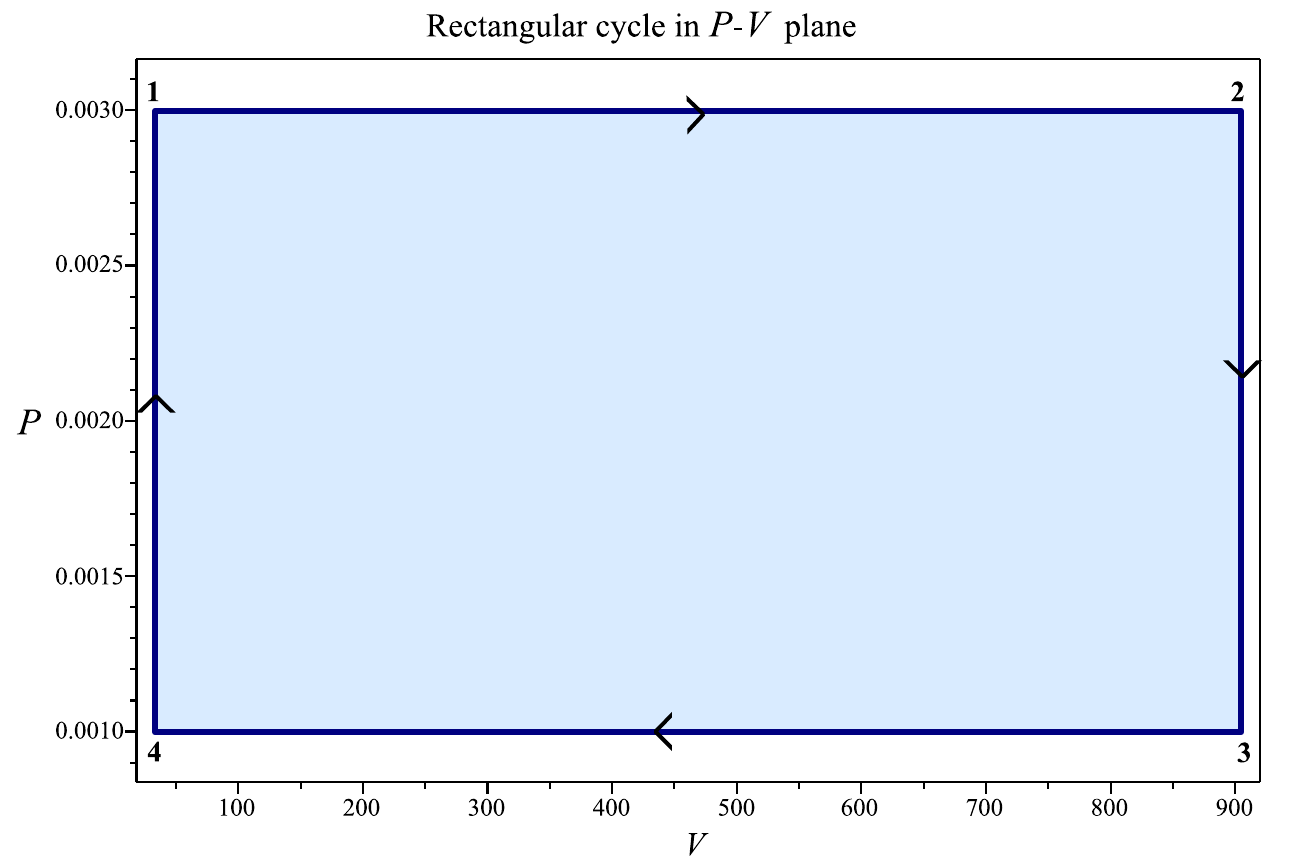}
\caption{Rectangular thermodynamic cycle in the $P$--$V$ plane for the holographic heat engine with $r_{+,1} = 2.0$, $r_{+,2} = 6.0$, $P_H = 0.003$, $P_L = 0.001$. The corners are labeled $1$--$4$, with arrows indicating the direction of the cycle. The BH absorbs heat $Q_H = M_2 - M_1 = 4.670$ along the high-pressure isobar $1 \to 2$, releases heat $Q_C = M_3 - M_4 = 2.927$ along the low-pressure isobar $3 \to 4$, and exchanges no $PdV$ work along the isochoric legs. The shaded area equals the net work $W = 1.743$.}
\label{fig:HE_cycle}
\end{figure}

Along the high-pressure isobar $1 \to 2$, the BH absorbs heat $Q_H = M_2 - M_1$, where $M_i = M(r_{+,i}, P_H)$ denotes the enthalpy at the corresponding corner. Along the low-pressure isobar $3 \to 4$, the BH releases heat $Q_C = M_3 - M_4$, with $M_3 = M(r_{+,2}, P_L)$ and $M_4 = M(r_{+,1}, P_L)$. The isochoric legs $2 \to 3$ and $4 \to 1$ involve no $PdV$ work since the volume is held fixed. The net work extracted by the engine is
\begin{equation}
W = Q_H - Q_C = (M_2 - M_1) - (M_3 - M_4) = (P_H - P_L)(V_2 - V_1)\,,
\label{eq:work}
\end{equation}
where the last equality follows from the fact that $V_i = 4\pi(r_{+,i}^3 + g^3)/3$ depends only on $r_{+,i}$ and $g$, not on $P$, so the rectangular cycle encloses a region of area $(P_H - P_L)(V_2 - V_1)$ in the $P$--$V$ plane. The engine efficiency is
\begin{equation}
\eta = \frac{W}{Q_H} = 1 - \frac{Q_C}{Q_H} = 1 - \frac{M(r_{+,2}, P_L) - M(r_{+,1}, P_L)}{M(r_{+,2}, P_H) - M(r_{+,1}, P_H)}\,.
\label{eq:eta}
\end{equation}
This expression is bounded above by the Carnot efficiency $\eta_C = 1 - T_C/T_H$, where $T_H = \max\{T_1, T_2, T_3, T_4\}$ and $T_C = \min\{T_1, T_2, T_3, T_4\}$ are the highest and lowest temperatures attained around the cycle. In the limit of large cycles ($r_{+,2} \to \infty$), the enthalpy is dominated by the $PV$ term and the efficiency approaches $\eta \to 1 - P_L/P_H$, which for our choice of pressures gives $\eta_{\rm max} = 1 - P_L/P_H = 2/3$.

For the numerical analysis we adopt the cycle parameters $r_{+,1} = 2.0$, $P_H = 0.003$, $P_L = 0.001$, and vary $r_{+,2}$ from $3$ to $13$ at the default BH parameters $g = 0.3$, $Q = 0.3$, $a = 0.1$, $\beta = 0.5$. These choices place $P_H$ slightly above the critical pressure $P_c \approx 0.0026$ from Sec.~\ref{isec4sub1} and $P_L$ well below it, so the cycle straddles the phase transition region. For the fixed cycle with $r_{+,2} = 6.0$, the thermodynamic volumes at the two corners are $V_1 \approx 33.62$ and $V_2 \approx 904.89$, and the corner temperatures are
\begin{equation}
T_1 = 0.0563\,,\quad T_2 = 0.0490\,,\quad T_3 = 0.0250\,,\quad T_4 = 0.0483\,.
\label{eq:corner_temps}
\end{equation}
The hottest corner is $T_1$ (small BH at high $P$) and the coldest is $T_3$ (large BH at low $P$). The enthalpy values $M_1 = 1.374$, $M_2 = 6.044$, $M_3 = 4.234$, $M_4 = 1.307$ yield $Q_H = 4.670$, $Q_C = 2.927$, and $W = 1.743$.

\subsection{Efficiency and parameter dependence} \label{isec6sub2}

The engine efficiency $\eta$ and the corresponding Carnot bound $\eta_C$ are plotted as functions of $r_{+,2}$ in Fig.~\ref{fig:HE_eta_default} for the default parameters. Both quantities increase monotonically with the cycle size $r_{+,2}$, reflecting the growth of the enclosed area in the $P$--$V$ plane. The actual efficiency $\eta$ (solid navy) starts at $\eta \approx 0.20$ for $r_{+,2} = 3$ and rises to $\eta \approx 0.56$ at $r_{+,2} = 13$, while the Carnot bound $\eta_C$ (dashed red) starts at $\eta_C \approx 0.40$ and reaches $\eta_C \approx 0.62$. The gap $\eta_C - \eta$ narrows as $r_{+,2}$ grows, indicating that larger cycles operate closer to the reversible limit. This is expected: for large $r_{+,2}$ the $PV$ term dominates the enthalpy at corner $2$, making $Q_H \approx (P_H - P_L) V_2$ and $\eta \approx 1 - P_L/P_H$, which coincides with the Carnot bound when the temperature spread becomes negligible compared to the overall scale. Both curves approach the asymptotic value $2/3$ from below.

\begin{figure}[ht!]
\centering
\includegraphics[width=0.85\textwidth]{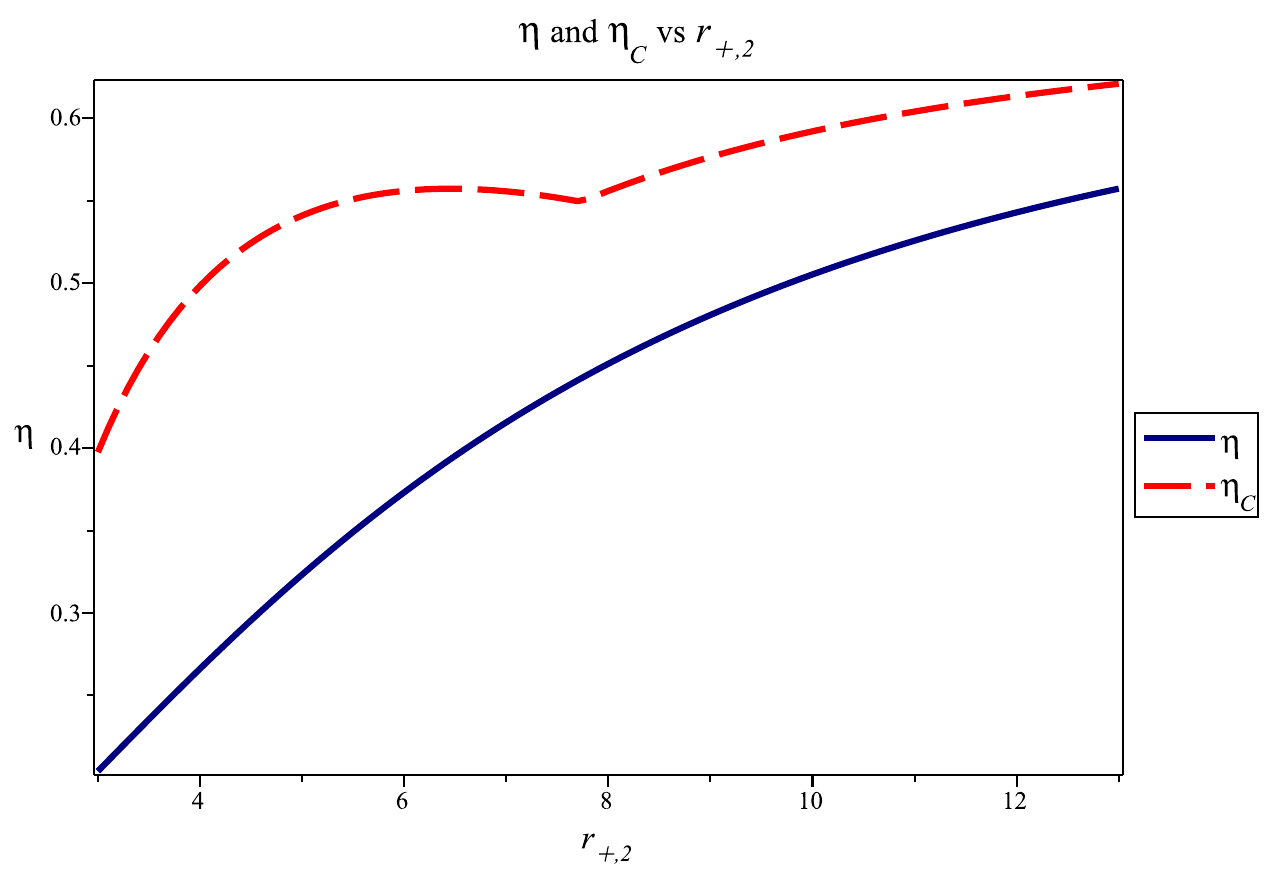}
\caption{Engine efficiency $\eta$ (solid navy) and Carnot bound $\eta_C$ (dashed red) as functions of $r_{+,2}$ for the rectangular cycle with $r_{+,1} = 2.0$, $P_H = 0.003$, $P_L = 0.001$ at the default BH parameters $g = 0.3$, $Q = 0.3$, $a = 0.1$, $\beta = 0.5$. Both increase monotonically with cycle size. The efficiency starts at $\eta \approx 0.20$ for $r_{+,2} = 3$ and reaches $\eta \approx 0.56$ at $r_{+,2} = 13$, remaining strictly below $\eta_C$ throughout. Both approach $1 - P_L/P_H = 2/3$ from below in the large-cycle limit.}
\label{fig:HE_eta_default}
\end{figure}

The effect of individual BH parameters on the efficiency is displayed in the four panels of Fig.~\ref{fig:HE_params}. Notably, the Hayward parameter $g$ (Fig.~\ref{fig:HE_g}) and the electric charge $Q$ (Fig.~\ref{fig:HE_Q}) have very weak effects on $\eta$: the four curves in each panel are nearly indistinguishable, overlapping within a band of width $\Delta\eta \lesssim 0.01$ across the entire range of $r_{+,2}$. The physical reason is that $g$ enters the thermodynamic volume as $V = 4\pi(r_+^3 + g^3)/3$ and the mass as $M \propto (r_+^3 + g^3)/r_+^2$, so its contribution to $Q_H = M_2 - M_1$ is suppressed by $g^3/r_{+,2}^3$ at the large-radius corner. Similarly, $Q$ enters via $Q^2/r_+^2$, which is small at both corners of the cycle (particularly at $r_{+,2} \gg 1$). The ratio $Q_C/Q_H$ is therefore nearly independent of $g$ and $Q$ for the chosen cycle geometry, and the efficiency is controlled primarily by the pressure ratio $P_L/P_H$ and the volume ratio $V_1/V_2$.

The CS parameter $a$ (Fig.~\ref{fig:HE_a}) has a clearly visible effect. Increasing $a$ from $0$ (navy) to $0.20$ (orange) raises $\eta$ by approximately $0.02$--$0.04$ across the full range of $r_{+,2}$, with the largest separation occurring at intermediate cycle sizes ($r_{+,2} \sim 5$--$8$). The ordering is monotonic: $a = 0.20$ yields the highest efficiency, followed by $a = 0.10$, $a = 0.05$, and $a = 0$. This behavior has a direct explanation: the CS parameter appears as the factor $(1-a)$ in the dominant contribution to the mass function. Increasing $a$ uniformly reduces all four corner masses by the same fraction, which lowers both $Q_H$ and $Q_C$. However, the work $W = (P_H - P_L)(V_2 - V_1)$ is independent of $a$ since the thermodynamic volume does not depend on the CS parameter. The ratio $W/Q_H$ therefore increases with $a$, yielding higher efficiency. In the language of the engine, the CS tension effectively lightens the working substance without changing the cycle geometry, making it easier to extract the same amount of work from a smaller heat input.

The PFDM parameter $\beta$ (Fig.~\ref{fig:HE_beta}) produces the most prominent separation among the four panels, and the ordering is opposite to that of $a$: $\beta = 0$ (navy) gives the highest efficiency, while $\beta = 1.0$ (orange) gives the lowest. The difference reaches $\Delta\eta \approx 0.05$ at $r_{+,2} \sim 5$ and diminishes at large $r_{+,2}$ as all curves converge toward $2/3$. The logarithmic PFDM contribution $\beta \ln(r_+/|\beta|)/r_+$ to the mass grows with $r_+$ (since $\ln r_+$ increases), so it preferentially increases $M_2$ and $M_3$ relative to $M_1$ and $M_4$. This raises $Q_H = M_2 - M_1$ more than it raises $W$, reducing the ratio $\eta = W/Q_H$. The PFDM therefore acts as a drag on the engine: it adds gravitational enthalpy at the expansion stage that must be supplied as heat input but does not contribute to the $PdV$ work output.

\begin{figure}[ht!]
\centering
\begin{subfigure}[b]{0.48\textwidth}
    \centering
    \includegraphics[width=\textwidth]{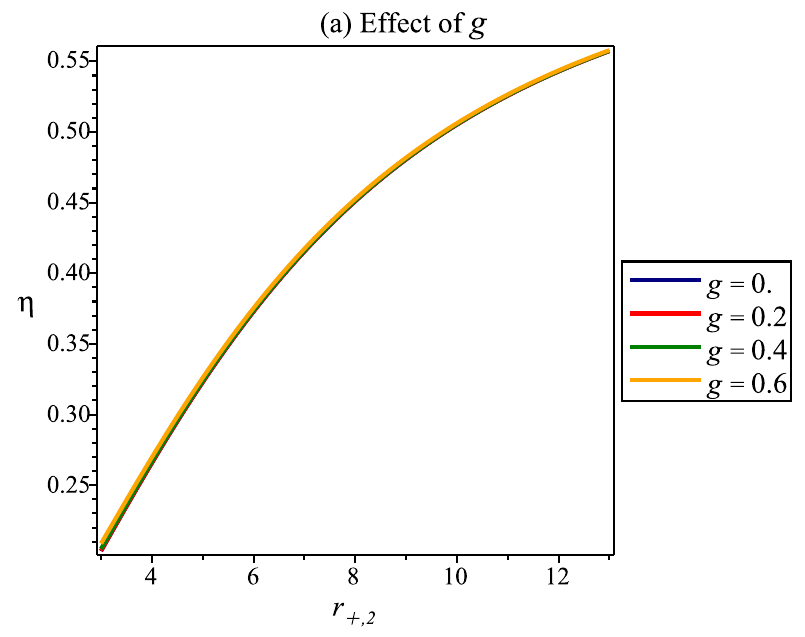}
    \caption{Effect of $g$ at $Q = 0.3$, $a = 0.1$, $\beta = 0.5$.}
    \label{fig:HE_g}
\end{subfigure}
\hfill
\begin{subfigure}[b]{0.48\textwidth}
    \centering
    \includegraphics[width=\textwidth]{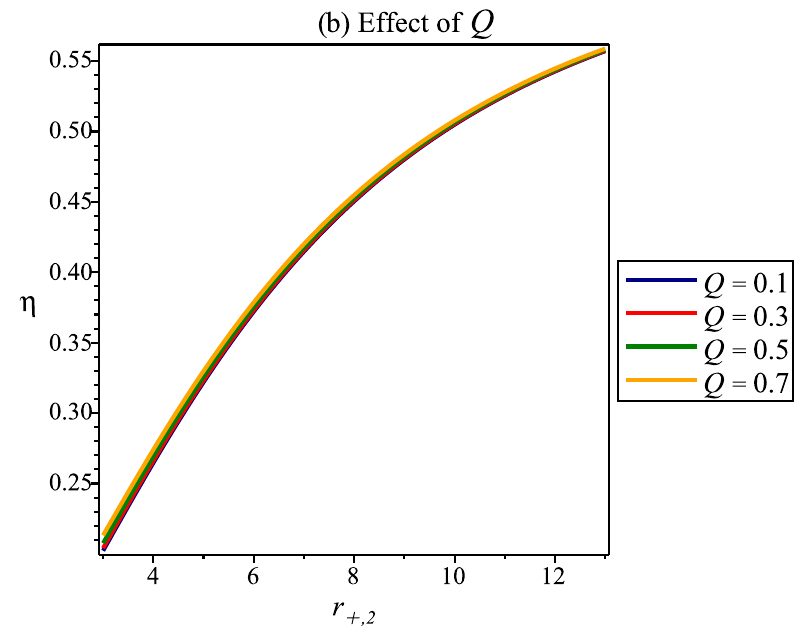}
    \caption{Effect of $Q$ at $g = 0.3$, $a = 0.1$, $\beta = 0.5$.}
    \label{fig:HE_Q}
\end{subfigure}

\vspace{0.5cm}

\begin{subfigure}[b]{0.48\textwidth}
    \centering
    \includegraphics[width=\textwidth]{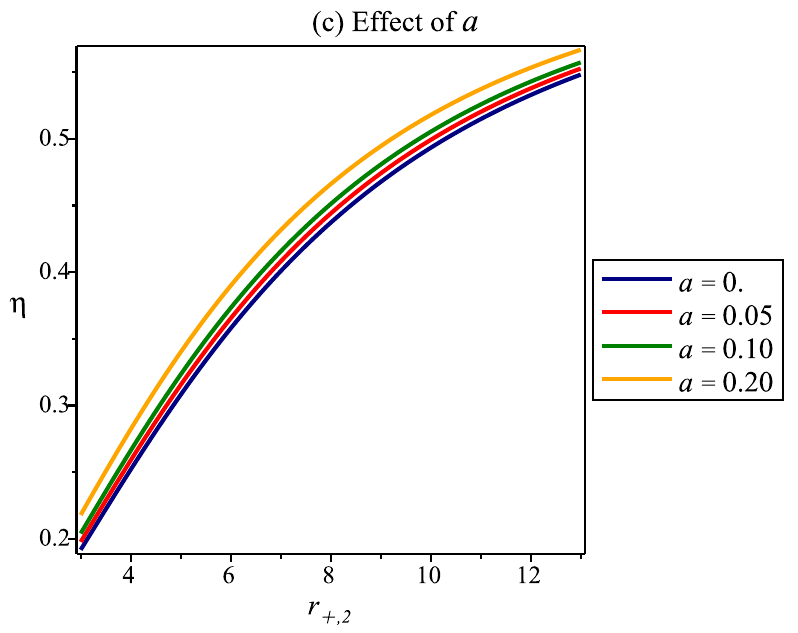}
    \caption{Effect of $a$ at $g = 0.3$, $Q = 0.3$, $\beta = 0.5$.}
    \label{fig:HE_a}
\end{subfigure}
\hfill
\begin{subfigure}[b]{0.48\textwidth}
    \centering
    \includegraphics[width=\textwidth]{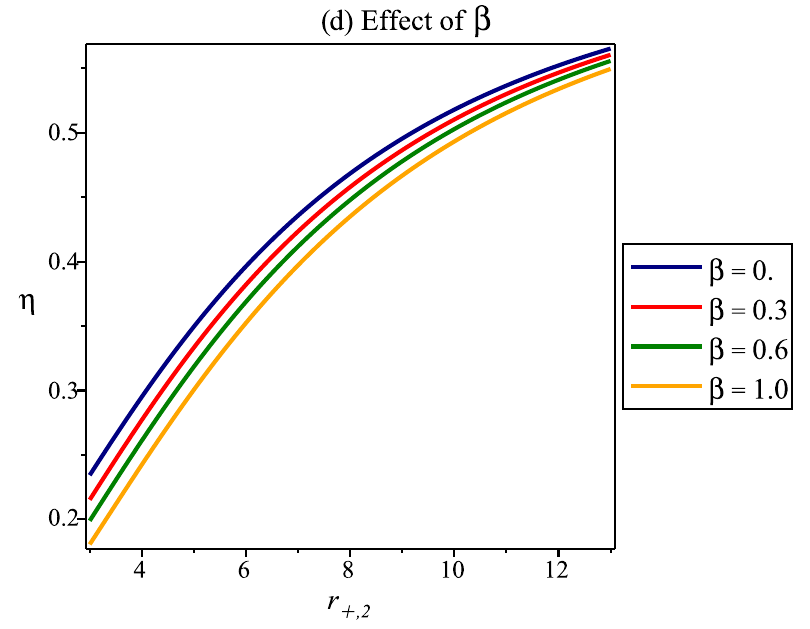}
    \caption{Effect of $\beta$ at $g = 0.3$, $Q = 0.3$, $a = 0.1$.}
    \label{fig:HE_beta}
\end{subfigure}
\caption{Effect of individual parameters on the heat engine efficiency $\eta(r_{+,2})$ for the rectangular cycle with $r_{+,1} = 2.0$, $P_H = 0.003$, $P_L = 0.001$. (a)~The Hayward parameter $g$ and (b)~the electric charge $Q$ have very weak effects on $\eta$, with all four curves nearly overlapping. (c)~The CS parameter $a$ has a clearly visible monotonic effect: increasing $a$ raises $\eta$ by $\sim\!0.02$--$0.04$. (d)~The PFDM parameter $\beta$ has the strongest effect with opposite sign: increasing $\beta$ lowers $\eta$ by up to $\sim\!0.05$ at intermediate $r_{+,2}$.}
\label{fig:HE_params}
\end{figure}

\subsection{Carnot benchmarking} \label{isec6sub3}

The ratio $\eta/\eta_C$ provides a dimensionless measure of how close the engine operates to the ideal Carnot limit. This benchmarking scheme, introduced by Johnson~\cite{Johnson2014HeatEngine}, is particularly informative for comparing different BH families, since it factors out the overall temperature scale and isolates the role of the equation of state.

\begin{figure}[ht!]
\centering
\includegraphics[width=0.85\textwidth]{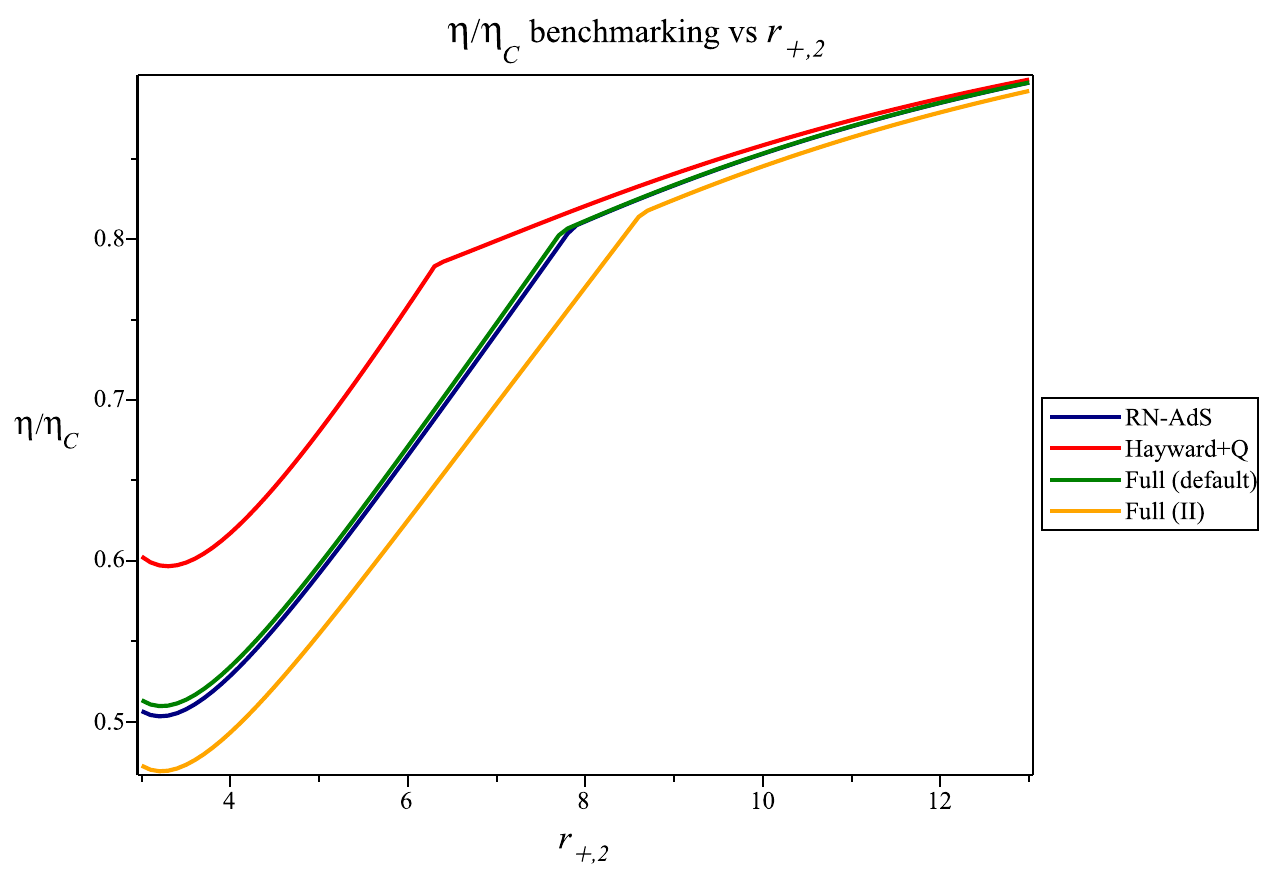}
\caption{Carnot benchmarking ratio $\eta/\eta_C$ as a function of $r_{+,2}$ for four BH configurations: RN-AdS (navy), Hayward+$Q$ (red), full model at default parameters (green), and full model at $g = 0.25$, $Q = 0.5$, $a = 0.1$, $\beta = 0.8$ (orange). The Hayward+$Q$ configuration has the highest benchmarking ratio at small $r_{+,2}$ ($\approx 0.60$), while Full~(II) with $\beta = 0.8$ has the lowest ($\approx 0.47$). All curves increase monotonically and converge near $\eta/\eta_C \approx 0.90$ for large cycles.}
\label{fig:HE_benchmark}
\end{figure}

The benchmarking ratio as a function of $r_{+,2}$ is displayed in Fig.~\ref{fig:HE_benchmark} for four configurations: RN-AdS (navy), Hayward+$Q$ without CS or PFDM (red), the full model at default parameters (green), and the full model at $g = 0.25$, $Q = 0.5$, $a = 0.1$, $\beta = 0.8$ (orange). All configurations satisfy $\eta/\eta_C < 1$ as required by the second law, and the ratio increases monotonically with $r_{+,2}$, approaching $\sim 0.90$ for large cycles. The most notable feature is the hierarchy at small cycle sizes. The Hayward+$Q$ configuration (red) has the highest benchmarking ratio, starting at $\eta/\eta_C \approx 0.60$ for $r_{+,2} = 3$, while the RN-AdS (navy) and full model (green) start at $\approx 0.50$--$0.52$, and the Full~(II) configuration with $\beta = 0.8$ (orange) starts lowest at $\approx 0.47$.

The superior performance of the Hayward+$Q$ engine (without CS or PFDM) can be traced to its smaller Carnot bound: from Table~\ref{tab:HE_efficiency}, $\eta_C = 0.500$ for Hayward+$Q$ compared to $0.556$ for the full model, yet $\eta$ is comparable ($0.379$ versus $0.373$). The Hayward+$Q$ system has a smaller temperature spread around the cycle because the PFDM logarithmic potential is absent, and this tighter temperature range brings $\eta$ closer to $\eta_C$ even though the absolute efficiency is similar. Adding PFDM (green and orange curves) increases the temperature spread at the large-radius corner and depresses the benchmarking ratio.

At large $r_{+,2}$, the four curves converge because the cycle geometry dominates over the matter content: the enthalpy at corner $2$ becomes $M_2 \approx (4\pi/3) P_H r_{+,2}^3$ regardless of $g$, $Q$, $a$, or $\beta$, and all engines approach the same asymptotic ratio.

\begin{table}[ht!]
\centering
\caption{Heat engine efficiency for six configurations at the fixed rectangular cycle $r_{+,1} = 2.0$, $r_{+,2} = 6.0$, $P_H = 0.003$, $P_L = 0.001$. The work output $W = 1.743$ is common to all configurations. The CS parameter improves both $\eta$ and $\eta/\eta_C$, while the PFDM parameter degrades them.}
\label{tab:HE_efficiency}
\vspace{0.3cm}
\renewcommand{\arraystretch}{1.6}
\setlength{\tabcolsep}{8pt}
\begin{tabular}{lcccccccc}
\hline
\rowcolor{orange!50}
Configuration & $g$ & $Q$ & $a$ & $\beta$ & $\eta$ & $\eta_C$ & $\eta/\eta_C$ & $W$ \\
\hline
RN-AdS ($g = 0$) & 0 & 0.3 & 0.1 & 0.5 & 0.3729 & 0.5602 & 0.6656 & 1.743 \\
Hayward+$Q$ & 0.3 & 0.3 & 0 & 0 & 0.3792 & 0.5000 & 0.7584 & 1.743 \\
Hayward+$Q$+CS & 0.3 & 0.3 & 0.1 & 0 & 0.3964 & 0.5010 & 0.7911 & 1.743 \\
Full model (default) & 0.3 & 0.3 & 0.1 & 0.5 & 0.3732 & 0.5560 & 0.6711 & 1.743 \\
Full model (II) & 0.25 & 0.5 & 0.1 & 0.8 & 0.3623 & 0.5795 & 0.6252 & 1.743 \\
Full model (III) & 0.3 & 0.5 & 0.05 & 0.6 & 0.3632 & 0.5597 & 0.6489 & 1.743 \\
\hline
\end{tabular}
\end{table}

The numerical values for six representative configurations at the fixed cycle $r_{+,1} = 2.0$, $r_{+,2} = 6.0$ are collected in Table~\ref{tab:HE_efficiency}. The work output $W = 1.743$ is the same for all configurations, confirming that $W = (P_H - P_L)(V_2 - V_1)$ depends only on the cycle geometry. The efficiency ranges from $\eta = 0.362$ (Full model~II, with $\beta = 0.8$) to $\eta = 0.396$ (Hayward+$Q$+CS), a variation of about $9\%$. The benchmarking ratio ranges from $\eta/\eta_C = 0.625$ to $0.791$, a spread of $27\%$, showing that $\eta/\eta_C$ is a more sensitive discriminator of the matter content than $\eta$ alone. The CS parameter consistently improves the benchmarking ratio (Hayward+$Q$ versus Hayward+$Q$+CS: $0.758 \to 0.791$), while the PFDM parameter degrades it (Hayward+$Q$+CS versus the full model: $0.791 \to 0.671$).

\section{Conclusions} \label{isec7}

We have constructed and analyzed the charged Hayward-AdS BH in the presence of a cloud of strings and perfect fluid dark matter, investigating its extended thermodynamic phase structure, JT expansion, and holographic heat engine performance. The solution is characterized by four independent parameters beyond the mass: the Hayward regularity parameter $g$, the electric charge $Q$, the CS parameter $a$, and the PFDM parameter $\beta$. The metric function~\eqref{eq:fmetric} interpolates between a regular dS core at $r \to 0$ and an asymptotically AdS geometry at $r \to \infty$, with the singularity resolution ensured by the Hayward factor $(r^3 + g^3)^{-1}$ for all $g > 0$. The main results are summarized below.

The horizon structure depends sensitively on the interplay among the four parameters. For the default set $g = 0.3$, $Q = 0.3$, $a = 0.1$, $\beta = 0.5$, the metric admits two horizons that merge at extremality. The Hayward parameter and the electric charge both shrink the range of masses admitting two horizons, while the CS parameter lowers the metric function uniformly through the factor $(1-a)$, and the PFDM logarithmic potential introduces qualitative changes in the small-$r$ regime. The extremal configurations have been mapped numerically and tabulated for six representative parameter choices.

In the extended phase space with $P = -\Lambda/(8\pi)$, the thermodynamic quantities --- Hawking temperature, entropy, thermodynamic volume, electric potential, and the conjugates to $a$, $\beta$, and $g$ --- have been derived and verified through the Smarr relation. A key subtlety is the distinction between the Hawking temperature $T_H = f'(r_+)/(4\pi)$ and the thermodynamic temperature $T_{\rm thermo} = (\partial M/\partial S)_P$, which differ by the factor $(r_+^3 + g^3)/r_+^3$ for $g \neq 0$. The Smarr identity~\eqref{eq:smarr} holds to machine precision ($\sim 10^{-10}$) once this distinction is properly accounted for.

The $P$--$V$ criticality analysis reveals a vdW-like first-order phase transition between small and large BH phases, with the critical point at $r_c = 4.130$, $T_c = 0.0410$, $P_c = 0.0026$ for the default parameters. The critical exponents $\alpha = 0$, $\tilde\beta = 1/2$, $\gamma = 1$, $\delta = 3$ confirm the mean-field universality class, identical to the vdW fluid and the RN-AdS BH. The GFE exhibits the characteristic swallowtail structure below $P_c$, degenerating to a cusp at $P = P_c$ and becoming smooth above it. The compressibility ratio $\rho_c \approx 0.524$ exceeds the vdW value $3/8 = 0.375$, with the PFDM contribution shifting the system further from vdW universality (increasing $\rho_c$ from $\approx 0.49$ without PFDM to $\approx 0.53$ with it).

The JT expansion analysis yields inversion curves that divide the $T_i$--$P$ plane into cooling ($\mu_{\rm JT} > 0$) and heating ($\mu_{\rm JT} < 0$) regions. The minimum inversion temperature satisfies $T_i^{\rm min}/T_c \approx 0.247$, which is roughly half the RN-AdS value ($\approx 0.5$) and one-third the value for ordinary fluids ($\approx 0.75$). This suppression originates from the combined effect of the Hayward volume correction $g^3$, the PFDM logarithmic potential, and the CS tension, all of which shift $T_c$ upward relative to $T_i^{\rm min}$ and compress the cooling window in temperature space. The parameter dependence of the inversion curves reveals a clear physical hierarchy: the Hayward parameter and the electric charge contract the cooling region (the $g = 0.6$ curve terminates at $P \approx 0.03$, while $Q = 0.7$ confines the curve to $P \lesssim 0.01$); the PFDM parameter expands it (the $\beta = 1.0$ curve extends to $T_i \approx 0.33$, compared to $T_i \approx 0.11$ for $\beta = 0$); and the CS parameter reshapes it non-monotonically, with intermediate $a$ values broadening the pressure range while reducing the peak temperature. The isenthalpic curves confirm that a BH undergoing constant-$M$ depressurization cools in the region below the inversion curve and heats above it, with larger $M$ extending the curves to higher temperatures.

The holographic heat engine operates through a rectangular cycle in the $P$--$V$ plane with $r_{+,1} = 2.0$, $P_H = 0.003$, $P_L = 0.001$, straddling the critical pressure. The engine efficiency $\eta$ increases monotonically with the cycle size $r_{+,2}$ and approaches $1 - P_L/P_H = 2/3$ from below. The parameter dependence reveals that the Hayward parameter and the electric charge have negligible effects on the efficiency (all curves overlap within $\Delta\eta \lesssim 0.01$), while the CS parameter $a$ improves it (raising $\eta$ by $\sim 0.02$--$0.04$ at intermediate $r_{+,2}$) and the PFDM parameter $\beta$ degrades it (lowering $\eta$ by up to $\sim 0.05$). For the fixed cycle at $r_{+,2} = 6.0$, the efficiency ranges from $\eta = 0.362$ (Full model~II, $\beta = 0.8$) to $\eta = 0.396$ (Hayward+$Q$+CS), with the Carnot benchmarking ratio spanning $\eta/\eta_C = 0.625$ to $0.791$. The CS parameter consistently improves the benchmarking ratio because it reduces the BH mass at fixed volume, effectively lightening the working substance. The PFDM degrades the ratio because the logarithmic potential adds gravitational enthalpy at the expansion stage without contributing to the $PdV$ work.

Several directions for future work suggest themselves. On the observational side, it would be interesting to study the BH shadow and the photon sphere for this geometry, which could connect the thermodynamic parameters to observational signatures from the EHT~\cite{EHT2019}. The quasinormal mode spectrum and the greybody factors would provide additional probes of the parameter space, particularly since the Hayward core and the PFDM potential modify the effective potential in distinct ways. On the theoretical side, the holographic interpretation of the CS and PFDM parameters in the dual CFT deserves further exploration, especially regarding the role of the PFDM logarithmic potential in the entanglement structure of the boundary theory. The extension to rotating Hayward-AdS geometries with CS and PFDM, though technically demanding, would allow the study of superradiance and the weak gravity conjecture in the presence of regular cores. Finally, the application of the restricted phase space formalism~\cite{RestrictedPhaseSpace2022}, which avoids the variation of $\Lambda$, could provide a complementary perspective on the phase transition structure reported here.

\section*{Acknowledgments}

F.A. gratefully acknowledges the Inter University Centre for Astronomy and Astrophysics (IUCAA), Pune, India, for the opportunity to serve as a visiting associate. \.{I}.~S. expresses his thanks to T\"{U}B\.{I}TAK, ANKOS, and SCOAP3 for their financial support. He further recognizes the backing of COST Actions CA22113, CA21106, CA23130, CA21136, and CA23115, which have played a crucial role in strengthening networking activities.

\section*{Data Availability Statement}

In this study, no new data was generated or analyzed.
\bibliographystyle{unsrtnat} 
\bibliography{ref}

}

\end{document}